\documentclass[aos,MSNbibl,seceqn,nameyear,dvips]{arximspdf}

\doi{10.1214/11-AOS917}
\volume{39}
\issue{5}
\pubyear{2011}
\firstpage{2740}
\lastpage{2765}

\makeatletter

\newcommand{\BF}{\operatorname{BF}}
\newcommand{\gbf}{g\mathrm{BF}}
\newcommand{\bgvn}{_{\gamma:N}}

\newtheorem{theorem}{Theorem}[section]
\newtheorem{lemma}{Lemma}[section]

\newproclaim{remark}{Remark}[section]

\newcommand{\plim}{\mathop{\operatorname{plim}}}
\newcommand{\bbm}{\mathbf}
\newcommand{\bbmm}{\bolds}

\makeatother

\begin{document}
\begin{frontmatter}

\title{Fully Bayes factors with a generalized $\bolds{g}$-prior}
\runtitle{Fully Bayes factors}

\begin{aug}
\author[A]{\fnms{Yuzo} \snm{Maruyama}\corref{}\ead[label=e1]{maruyama@csis.u-tokyo.ac.jp}}
\and
\author[B]{\fnms{Edward I.} \snm{George}\ead[label=e2]{edgeorge@wharton.upenn.edu}}
\runauthor{Y. Maruyama and E. I. George}
\affiliation{University of Tokyo and University of Pennsylvania}
\address[A]{Center for Spatial Information Science\\
University of Tokyo\\
5-1-5 Kashiwanoha, Kashiwa-shi\\
Chiba, 277-8568\\
Japan\\
\printead{e1}} 
\address[B]{Department of Statistics\\
University of Pennsylvania\\
400 Jon M. Huntsman Hall\\
3730 Walnut Street\\
Philadelphia, Pennsylvania 19104-6302\\
USA\\
\printead{e2}}
\end{aug}

\received{\smonth{9} \syear{2010}}
\revised{\smonth{8} \syear{2011}}

%
\begin{abstract}
For the normal linear model variable selection problem,
we propose selection criteria based on a fully Bayes formulation
with a generalization of Zellner's $g$-prior which allows for $p > n$.
A special case of the prior formulation is seen
to yield tractable closed forms for marginal densities
and Bayes factors which reveal new model
evaluation characteristics of potential interest.
\end{abstract}

%

\end{frontmatter}

\section{Introduction}
\label{secintro}
Suppose the normal linear regression model is used to relate $y$ to the
potential predictors $x_1,\ldots, x_p$,
%
%
\begin{equation} \label{full-model}
\bbm{y} \sim N_n(\alpha\bbm{1}_n+\bbm{X}_F\bbmm{\beta}_F,\sigma^2 \bbm{I}_n),
\end{equation}
where $\alpha$ is an unknown intercept parameter,
$\bbm{1}_n$ is an $n \times1$ vector each component of which is one,
$\bbm{X}_F=(\bbm{x}_1,\ldots, \bbm{x}_p)$ is an $n \times p$ design matrix,
$\bbmm{\beta}_F$ is a $p \times1$ vector of unknown
regression coefficients, $\bbm{I}_n$ is an $n\times n$ identity matrix
and $\sigma^2$ is an unknown positive scalar.
(The subscript $F$ denotes the full model.)
We assume that the columns of $\bbm{X}_F$ have been standardized so
that for
$1 \leq i \leq p$, $\bbm{x}'_i\bbm{1}_n = 0$ and
$\bbm{x}'_i\bbm{x}_i/n =1$.

We shall be particularly interested in the variable selection
problem where we would like to select an unknown
subset of the important predictors.
It will be convenient throughout to index each of these $2^p$ possible
subset choices by the vector
\[
\bbmm{\gamma} =(\gamma_1,\ldots,\gamma_p)',
\]
where $\gamma_i=0$ or $1$.
We use $q_\gamma=\bbmm{\gamma}'\bbm{1}_p$ to denote the size of
the $\bbmm{\gamma}$th subset.
The problem then becomes that of selecting a submodel
of (\ref{full-model}) which has a density of the form
%
%
\begin{equation} \label{submodel-gamma}
p(\bbm{y}|\alpha, \bbmm{\beta}_\gamma, \sigma^2,\bbmm{\gamma})
=\phi_n(\bbm{y}; \alpha\bbm{1}_n+ \bbm{X}_\gamma\bbmm{\beta}_\gamma
,\sigma^2 \bbm{I}_n),
\end{equation}
where $\phi_n(\bbm{y}; \bbmm{\mu}, \bbmm{\Sigma})$
denotes the $n$-variate normal density with mean vector~$\bbmm{\mu}$ and
covariance matrix $\bbmm{\Sigma}$.
In (\ref{submodel-gamma}), $\bbm{X}_\gamma$ is the $n \times q_\gamma$
matrix whose columns correspond to the $\bbmm{\gamma}$th subset
of $x_1,\ldots,x_p$, and $\bbmm{\beta}_\gamma$
is a $q_\gamma\times1$ vector of unknown regression coefficients.
We assume throughout that $\bbm{X}_\gamma$ is of full rank denoted
\[
r_\gamma= \min\{q_\gamma, n-1 \}.
\]
Last, let $\mathcal{M}_\gamma$ denote the submodel given by (\ref
{submodel-gamma}).

A Bayesian approach to this problem entails the specification of
prior distributions on the models $\pi_\gamma=
\operatorname{Pr}(\mathcal{M}_\gamma)$, and on the parameters
$p(\alpha,\bbmm{\beta}_\gamma, \sigma^2)$ of each model.
For each such specification,
of key interest is the posterior probability of $\mathcal{M}_\gamma$
given $\bbm{y}$,
%
%
\begin{equation} \label{posterior-2}
\operatorname{Pr}(\mathcal{M}_\gamma|\bbm{y})
=\frac{\pi_\gamma m_\gamma(\bbm{y})}{\sum_\gamma\pi_\gamma
m_\gamma(\bbm{y})}
=\frac{\pi_\gamma\BF\bgvn}
{\sum_\gamma\pi_\gamma\BF\bgvn},
\end{equation}
where $m_\gamma(\bbm{y})$ is the marginal density of $\bbm{y}$ under
$\mathcal{M}_\gamma$.
In (\ref{posterior-2}), $ \BF\bgvn$
is the so-called ``null-based Bayes factor'' for comparing
each of $ \mathcal{M}_\gamma$ to the null model $\mathcal{M}_N$
which is defined as
\[
\BF\bgvn= \frac{m_\gamma(\bbm{y})}{m_N(\bbm{y})},
\]
where the null model $\mathcal{M}_N$ is given by
$ \bbm{y} \sim N_n(\alpha\bbm{1}_n,\sigma^2 \bbm{I}_n)$ and $ m_N(\bbm
{y})$ is the marginal density of $\bbm{y}$ under the null model.
For model selection, a popular strategy is to select the model for which
$\operatorname{Pr}(\mathcal{M}_\gamma|\bbm{y})$ or $ \pi_\gamma\BF\bgvn$
is largest.

Our main focus in this paper is to propose and study specifications for
the parameter prior
for each submodel $\mathcal{M}_\gamma$, which we will consider to be
of the form
%
%
\begin{eqnarray}\label{priorform}
p(\alpha,\bbmm{\beta}_\gamma, \sigma^2)&=&p(\alpha)p(\sigma^2)
p(\bbmm{\beta}_\gamma|\sigma^2)\nonumber\\[-8pt]\\[-8pt]
&=&
p(\alpha)p(\sigma^2)\int
p(\bbmm{\beta}_\gamma|\sigma^2,g)p(g)\,dg,\nonumber
\end{eqnarray}
where $g$ is a hyperparameter.
In Section \ref{secpriors}
we explicitly describe our choices of prior forms for (\ref{priorform}).
Our key innovation there will be to use a generalization of
%
%
\begin{equation}\label{g-prior}
p(\bbmm{\beta}_\gamma|\sigma^2,g)
= \phi_{q_\gamma}(\bbmm{\beta};\bbm{0}, g \sigma^2(\bbm{X}'_\gamma
\bbm{X}_\gamma)^{-1}),
\end{equation}
Zellner's (\citeyear{Zellner-1986}) $g$-prior,
a normal conjugate form which leads to tractable marginalization, for example,
see \citet{George-Foster-2000}, \citet{Fernandez-Ley-Steel-2001},
\citet{Liang-etal-2008}.
Under (\ref{g-prior}) and a flat prior on $\alpha$,
the marginal density of $\bbm{y}$ given $g$ and $\sigma^2$ under
$\mathcal{M}_\gamma$ is
given by
%
%
\begin{equation} \label{marginal-known-1}
m_\gamma(\bbm{y}|g,\sigma^2)
\propto\exp\biggl( \frac{g}{g+1}
\Bigl\{\max_{\alpha,\bbmm{\beta}_\gamma} \log p(\bbm{y}| \alpha,
\bbmm{\beta}_\gamma,\sigma^2)
-q_\gamma H(g) \Bigr\}\biggr),
\end{equation}
where $H(g)= (2g)^{-1}(g+1)\log(g+1)$,
a special case of the key relation in \citet{George-Foster-2000}.
As they point out, for particular values of~$g$, when~$\sigma^2$ is known,
the Bayesian strategy of choosing $\mathcal{M}_\gamma$
to maximize (\ref{marginal-known-1}) corresponds to common fixed
penalty selection criteria.
For example, setting $H(g)=2$, $\log n$ or $2\log p$ (independently of
$\bbm{y}$)
would correspond to AIC [\citet{Akaike-1974}], BIC [\citet
{Schwarz-1978}]
or RIC [\citet{Foster-George-1994}], BIC,
or RIC, respectively.
For a discussion of recommendations in the literature
for choosing a fixed $g$ depending on $p$ and/or $n$,
see Section~2.4 of \citet{Liang-etal-2008}.

Although the correspondences to fixed penalty criteria are interesting,
as a practical matter,
it is necessary to deal with the uncertainty about $g$ and $\sigma^2$
to obtain useful criteria.
For this purpose, \citet{George-Foster-2000} proposed selecting
the model
maximizing $m_\gamma(\bbm{y}|g,\sigma^2)$ based on an empirical Bayes estimate
of $g$ and the standard unbiased estimate of $\sigma^2$.
More recently, \citet{Cui-George-2008} proposed margining out $g$
with respect to a prior, and \citet{Liang-etal-2008} proposed margining
out $g$
and $\sigma^2$ with respect to priors.
It should be noted that the first paper to effectively use a prior integrating
out~$g$ was \citet{Zellner-Siow-1980}; they stated things in terms of
multivariate Cauchy densities, which can always be expressed as a
$g$-mixture of $g$-priors.
All of these strategies lead to criteria that can be seen
as adapting to the fixed penalty criterion which would be most suitable
for the data at hand.
In this paper, we shall similarly follow a fully Bayes approach,
but with a~generalization of the $g$-prior (\ref{g-prior})
and an extension of the considered class of priors on $g$.

After describing our prior forms in Section \ref{secpriors}
and then calculating the marginals and Bayes factors in Section \ref
{secmarginal+BF},
we ultimately obtain our proposed $g$-prior Bayes factor ($g$BF),
which is of the form (omitting the $\gamma$ subscripts for clarity)
%
%
\begin{equation}\label{gBF}
\gbf\bgvn=
\cases{
\displaystyle \biggl\{\frac{\bar{d}}{d_q}\biggr\}^{-q}
\frac{\{1- R^2+ d_q^2 \|\hat{\bbmm{\beta}}_{\mathrm{LS}}\|^2\}
^{-{1/4}-{q/2}}}
{C_{n,q}(1- R^2 )^{(n-q)/2-3/4}},
&\quad if $q < n-1$,\vspace*{2pt}\cr
\{\bar{d} \times\| \hat{\bbmm{\beta}}^{\mathrm{MP}}_{\mathrm{LS}}\|\}^{-n+1},
&\quad if $q \geq n-1$,}\hspace*{-28pt}
\end{equation}
where $C_{n,q} \equiv\frac{B(1/4,(n-q)/2-3/4)}{B(q/2+1/4,(n-q)/2-3/4)}$
using the Beta function $B(\cdot,\cdot)$,
$ R^2$ is the familiar $R$-squared statistic under $\mathcal{M}_\gamma$,
$\bar{d}$ and $d_r$ are, respectively, the geometric mean and minimum
of the singular values of $\bbm{X}_\gamma$, \mbox{$ \| \cdot\|$} is the
$L_2$ norm,
and finally, for the standardized response
$(\bbm{y}-\bar{y}\bbm{1}_n)/\| \bbm{y}-\bar{y}\bbm{1}_n \|$, $\hat
{\bbmm{\beta}}_{\mathrm{LS}}$
is the usual least squares estimator, and $ \hat{\bbmm{\beta}}^{\mathrm{MP}}_{\mathrm{LS}}$
is the least squares estimator using the Moore--Penrose inverse matrix.

Two immediately apparent features of (\ref{gBF}) should be noted.
First, in contrast to other fully Bayes factors for our selection problem,
$\gbf$ is a~closed form expression which allows for
interpretation
and straightforward calculation under\vadjust{\goodbreak} any model.
As will be seen in later sections, this transparency reveals that $\gbf$
not only rewards explained variation overall,
but also rewards variation explained by the larger principal components
of the design matrix.
Second, $\gbf$ can be applied to all models
even when the number of predictors $p$ exceeds the number of
observations $n$.
This includes $p > n$ which is of increasing interest.
This is not the case for (\ref{g-prior}) which requires $p \le n-1$
so that $\bbm{X}'_\gamma\bbm{X}_\gamma$ will be invertible for
all~$q_\gamma$,
(recall that $\bbm{X}_\gamma$ has dimension at most $n-1$
because its columns have been centered).
Note also that when $p > n-1$, penalized sum-of-squares criteria
such as AIC, BIC and RIC will be unavailable for all submodels.

The organization of this paper is as follows.
In Section \ref{secpriors}
we propose prior forms including a generalized $g$-prior with a
beta-prime prior for $g$.
In Section \ref{secmarginal+BF} we derive general Bayes factor
expressions, and propose
default hyperparameter settings which yield $\gbf$ above.
In Section \ref{sechyper} we discuss appealing consequences of our default
specifications.
In Section \ref{secafter} we describe conditional shrinkage estimation
with the generalized $g$-prior.
In Section \ref{secconsistency} we show that $\gbf$
is consistent for model selection as $n \to\infty$.
In Section \ref{secsim} we provide a simulation evaluation of $\gbf$
performance.

\section{A fully Bayes prior formulation}
\label{secpriors}
We now proceed to describe the prior components that form
$p(\alpha,\bbmm{\beta}_\gamma, \sigma^2)$ in (\ref{priorform}).
Throughout the remainder of the paper,
we will omit the subscript $\gamma$ for notational simplicity when
there is no ambiguity. However, it is important to remember throughout
that our formulations are to be applied to all of the $2^p$ possible
submodels in (\ref{submodel-gamma}).

\subsection{\texorpdfstring{A generalized $g$-prior for $\beta$}{A generalized g-prior for beta}}
To motivate our proposed generalization of Zellner's $g$-prior,
we begin with a reconsideration of the original $g$-prior (\ref{g-prior})
for the case $p \leq n-1$.
The covariance matrix of the $g$-prior, $g\sigma^2(\bbm{X}'\bbm{X})^{-1}$,
is proportional to the covariance matrix of the least squares estimator
$\hat{\bbmm{\beta}}_{\mathrm{LS}}$.
As a consequence of this choice, the marginal likelihood with respect to
the $g$-prior
appealingly becomes a function only of the residual sum-of-squares, RSS.

However, from the ``matrix conditioning'' viewpoint of
Casella (\citeyear{Casella-1980,Casella-1985})
which advocates more shrinkage on higher variance estimates,
the original $g$-prior may not be reasonable.
To see why, let us rotate the problem by the $q \times q$
orthogonal matrix
$\bbm{W}=(\bbm{w}_1,\ldots,\bbm{w}_q)$ which diagonalizes~$\bbm{X}' \bbm
{X}$ as
%
%
\begin{equation} \label{evd-f}
\bbm{W}'(\bbm{X}'\bbm{X})\bbm{W}=\bbm{D}^2,
\end{equation}
where $\bbm{D}=\operatorname{diag}(d_1,\ldots, d_q)$
with
%
%
\begin{equation}
d_1 \geq\cdots\geq d_q >0.
\end{equation}
Thus,
\[
\bbm{W}'\hat{\bbmm{\beta}}_{\mathrm{LS}} \sim N_q(\bbm{W}'\bbmm{\beta}, \sigma
^2\bbm{D}^{-2}).
\]
Applying the $g$-prior (\ref{g-prior}) to these rotated coordinates
would then induce the prior
\[
\bbm{W}'\bbmm{\beta} \sim N_q(\bbm{0}, g \sigma^2\bbm{D}^{-2}),
\]
which reveals the prior variances to be proportional to the sample
variances of the elements of
$\bbm{W}'\hat{\bbmm{\beta}}_{\mathrm{LS}}$. This contradicts
\citet{Casella-1980} who states, ``if the sampling information is
good, it is
reasonable to downweight the prior guess.''
To remedy this situation, we propose consideration of priors on $\bbmm
{\beta}$ for which
\[
\bbm{W}'\bbmm{\beta} \sim N_q(\bbm{0},\sigma^2 \bbmm{\Psi}_q),
\]
where the components of $\bbmm{\Psi}_q=\operatorname{diag}(\psi_1,\ldots
,\psi
_q)$ are in descending order, namely,
%
%
\begin{equation} \label{des-order}
\psi_1 \geq\cdots\geq\psi_q>0.
\end{equation}
Note that this would be satisfied for $\bbmm{\Psi}_q \propto\bbm
{I}_q$, a consequence of the common assumption of exchangeable $\bbmm
{\beta}$ components.

In fact, a slightly weaker ordering of the form
%
%
\begin{equation} \label{weak-des-order}
d_1^2\psi_1 \geq\cdots\geq d_q^2\psi_q >0
\end{equation}
would still be reasonable because
the resulting Bayes estimator of $\bbm{w}'_i\bbmm{\beta}$ would be of
the form
\[
(1+\{d_i^2\psi_i\}^{-1})^{-1}\bbm{w}'_i\hat{\bbmm{\beta}}_{\mathrm{LS}},
\]
so that under (\ref{weak-des-order}),
the components of $\bbm{W}'\hat{\bbmm{\beta}}_{\mathrm{LS}}$ with larger
variance would be shrunk more.
We note that the original $g$-prior (\ref{g-prior}),
for which $\psi_i =g d_i^{-2}$, satisfies only the extreme boundary of
(\ref{weak-des-order}), namely,
\[
d_1^2\psi_1 =\cdots= d_q^2\psi_q=g.
\]
This violates (\ref{des-order}) whenever $d_i > d_{i+1}$,
in which case
$\psi_i< \psi_{i+1}$.

An appealing general form for $\bbmm{\Psi}_q$ is
$ \bbmm{\Psi}_q(g,\bbmm{\nu})=\operatorname{diag}(\psi_1(g,\bbmm{\nu
}),\ldots, \psi_q(g,\bbmm{\nu}))$,
where
%
%
\begin{equation} \label{general-psi}
\psi_i(g,\bbmm{\nu})=(1/d_i^2)\{ \nu_i(1+g)-1 \},
\end{equation}
$ \bbmm{\nu}=(\nu_1,\ldots,\nu_q)'$ and $\nu_i \geq1$ for any $i$,
guaranteeing
$\psi_i(g,\bbmm{\nu}) >0$.
Note that $ \bbmm{\Psi}_q(g,\bbmm{\nu})$, like the original $g$-prior,
is controlled by a single hyperparameter $g>0$.
When $\nu_1 =\cdots=\nu_q=1$, $ \sigma^2\bbmm{\Psi}_q(g,\bbmm{\nu})
$ becomes
$g\sigma^2\bbm{D}^{-2}$, yielding
the covariance structure of the original $g$-prior.
Although (\ref{weak-des-order}) will be satisfied whenever
$ \nu_1 \geq\cdots\geq\nu_q \geq1$,
we shall ultimately be interested
in a particular design dependent choice defined in Section \ref{secdefaults}.
In summary, when $q \leq n-1$, we propose a generalized $g$-prior for
$\bbmm{\beta}$ of the form
%
%
\begin{equation}\label{gen-g-psmall}
p(\bbmm{\beta} | \sigma^2,g)
=\phi_q(\bbm{W}'\bbmm{\beta};\bbm{0},\sigma^2 \bbmm{\Psi}_q(g,\bbmm
{\nu})),
\end{equation}
where $ \nu_1 \geq\cdots\geq\nu_q \geq1$.\vadjust{\goodbreak}

When $ q > n-1$ and the rank of $\bbm{X}$ is $n-1$,
there exists a $q \times(n-1)$ matrix
$\bbm{W}=(\bbm{w}_1,\ldots,\bbm{w}_{n-1})$ which diagonalizes $\bbm
{X}'\bbm{X}$ as
%
%
\begin{equation} \label{evd-m}
\bbm{W}'(\bbm{X}'\bbm{X})\bbm{W}=\bbm{D}^2,
\end{equation}
where $\bbm{W}'\bbm{W}=\bbm{I}_{n-1}$ and $\bbm{D}=\operatorname{diag}(d_1,
d_2,\ldots, d_{n-1})$
with $ d_1 \geq d_2 \geq\cdots\geq d_{n-1} >0$.
For this case, we propose a generalized $g$-prior of the form
%
%
\begin{equation}
p(\bbmm{\beta}|\sigma^2,g)
=\phi_{n-1}(\bbm{W}'\bbmm{\beta};\bbm{0},\sigma^2 \bbmm{\Psi
}_{n-1}(g,\bbmm{\nu}))
p_\#(\bbm{W}'_\#\bbmm{\beta}),
\end{equation}
where
$\bbmm{\Psi}_{n-1}(g,\bbmm{\nu})=\operatorname{diag}(\psi_1,\ldots,\psi_{n-1})$
is again given by (\ref{general-psi})
and $\nu_1 \geq\cdots\geq\nu_{n-1}\geq1$. Here, $\bbm{W}_\# $ is
an arbitrary
matrix which makes the $q\times q$ matrix
$(\bbm{W}, \bbm{W}_\#)$ orthogonal, and $p_\#(\cdot)$ is an arbitrary
probability
density on $\bbm{W}'_\#\bbmm{\beta}$, respectively.
As will be seen, the choices of $\bbm{W}_\#$ and $p_\#$
have no effect on the selection criteria we obtain, thus we leave them
as arbitrary.\looseness=1

Combining the above two cases by letting
%
%
\begin{equation}
r=\min\{q, n-1\},
\end{equation}
our suggested generalized $g$-prior is of the form
%
%
\begin{eqnarray} \label{prior-beta-1}
p(\bbmm{\beta}|g,\sigma^2)
&=&\phi_r(\bbm{W}'\bbmm{\beta};\bbm{0},\sigma^2 \bbmm{\Psi}_r(g,\bbmm
{\nu})) \nonumber\\[-8pt]\\[-8pt]
&&{}\times
\cases{
1, &\quad if $q \leq n-1$, \cr
p_\#(\bbm{W}'_\# \bbmm{\beta}), &\quad if $q > n-1$,}\nonumber
\end{eqnarray}
where the $q\times r$ matrix $\bbm{W}$
satisfies both
$ \bbm{W}'\bbm{X}'\bbm{X}\bbm{W} =\operatorname{diag}(d_1^2,\ldots,d_r^2)$
and $\bbm{W}'\bbm{W}=\bbm{I}_{r}$,
and $ \bbmm{\Psi}_r(g,\bbmm{\nu})=\operatorname{diag}(\psi_1(g,\bbmm{\nu
}),\ldots,\psi_r(g,\bbmm{\nu}))$
with (\ref{general-psi}).
%
%
\begin{remark}
In (\ref{evd-f}) and (\ref{evd-m}), let
%
%
\begin{equation} \label{u}
\bbm{U}=(\bbm{u}_1,\ldots,\bbm{u}_r)=(\bbm{Xw}_1/d_1,\ldots,\bbm
{Xw}_r/d_r)=\bbm{XWD}^{-1}.
\end{equation}
Then $\bbm{U}'\bbm{U}=\bbm{I}_r$ and
%
%
\begin{equation} \label{svd-1}
\bbm{X}=\bbm{UDW}'=\sum_{i=1}^r d_i \bbm{u}_i \bbm{w}'_i.
\end{equation}
This is the nonnull part of the well-known singular value
decomposition (SVD).
The diagonal elements of $\bbm{D} = \operatorname{diag}(d_1,\ldots,d_r)$
are the singular values of $\bbm{X}$, and the columns of
$\bbm{U}=(\bbm{u}_1,\ldots,\bbm{u}_r)$ are the normalized principal components
of the column space of $\bbm{X}$.
Note that the components of the rotated vector $\bbm{W}'\bbmm{\beta}$
are the coefficients for the principal component regression of $\bbm
{y}$ on $\bbm{UD}$.
From the definition of $\bbm{W}$ and $\bbm{U}$ by (\ref{evd-f}), (\ref
{evd-m}) and~(\ref{u}),
the signs of $\bbm{u}_i\bbm{w}'_i$ are determinate
although the signs of $\bbm{w}_i$ and $\bbm{u}_i$
for $1 \leq i \leq r$ are indeterminate.
These indeterminacies can safely be ignored in our development.
\end{remark}

\subsection{\texorpdfstring{Priors for $g$, $\alpha$ and $\sigma^2$}{Priors for g, alpha and sigma^2}}
Turning to the prior for the hyperparameter~$g$, we propose
%
%
\begin{equation}\label{prior-g}
p(g) =
\frac{g^b(1+g)^{-a-b-2}}{B(a+1,b+1)}
I_{(0, \infty)}(g)
\end{equation}
with $a> -1$, $ b>-1$, a Pearson Type VI or
\textit{beta-prime} distribution under which $1/(1+g)$ has a Beta
distribution $\operatorname{Be}(a+1,b+1)$.
Choices for the hyperparameters $a$ and $b$ are discussed later.

Although \citet{Zellner-Siow-1980} did not explicitly use a $g$-prior
formulation with a
prior on $g$,
their recommendation of a multivariate Cauchy form for $ p(\bbmm{\beta
}|\sigma^2)$
implicitly corresponds to using a $g$-prior with an inverse Gamma prior
\[
(n/2)^{1/2}\{\Gamma(1/2)\}^{-1} g^{-3/2} e^{-n/(2g)}
\]
on $g$. Both \citet{Cui-George-2008} and \citet
{Liang-etal-2008} proposed
using $g$-priors with priors of the form
%
%
\begin{equation} \label{prior-liang}
p(g)=(a+1)^{-1}(1+g)^{-a-2},
\end{equation}
the subclass of (\ref{prior-g}) with $b=0$.
Cases for which $b= O(n)$ will be of interest to us in what follows.

For the parameter $\alpha$ and $\sigma^2$, we use the location
invariant flat prior
%
%
\begin{equation} \label{prior-alpha}
p(\alpha)=I_{(-\infty,\infty)}(\alpha)
\end{equation}
and the scale invariant prior
%
%
\begin{equation} \label{prior-sigma}
p(\sigma^2)=(\sigma^2)^{-1}I_{(0,\infty)}(\sigma^2),
\end{equation}
respectively. Because $\alpha$ and $\sigma^2$ appear in every model,
the use of these improper priors for Bayesian model selection is
formally justified by \citet{Ber-Per-Var-1998}.

We note in passing that for the estimation of a
multivariate normal mean, priors equivalent to
(\ref{gen-g-psmall}), (\ref{prior-g}), (\ref{prior-alpha}) and (\ref
{prior-sigma})
have been considered by \citet{Strawderman-1971} and extended by
\citet{Maru-Straw-2005}.

\section{Marginal densities and Bayes factors}
\label{secmarginal+BF}
\subsection{General forms}
\label{secgeneral-forms}
The marginal densities of $\bbm{y}$
under $\mathcal{M}_\gamma(\mbox{$\neq$}\mathcal{M}_N)$\break and~$\mathcal{M}_N$
are, by definition,
%
%
\begin{eqnarray}\label{full-marginal}
m_\gamma(\bbm{y})&=&
\int_{-\infty}^{\infty} \int_{R^{q}}
\int_{0}^{\infty}
p(\bbm{y}|\alpha,\bbmm{\beta}_\gamma, \sigma^2) p(\alpha,\bbmm{\beta
}_\gamma,\sigma^2)
\,d \alpha \,d \bbmm{\beta}_\gamma \,d\sigma^2, \nonumber\\[-8pt]\\[-8pt]
m_N(\bbm{y})&=&
\int_{-\infty}^{\infty}
\int_{0}^{\infty}
p(\bbm{y}|\alpha, \sigma^2) p(\alpha,\sigma^2)
\,d \alpha \,d\sigma^2,\nonumber
\end{eqnarray}
respectively. Under the priors
\[
p(\alpha,\bbmm{\beta}_\gamma, \sigma^2)
=p(\alpha)p(\sigma^2)\int_0^\infty p(\bbmm{\beta}_\gamma|\sigma
^2,g)p(g)\,dg \qquad\mbox{for }\mathcal{M}_\gamma(\mbox{$\neq$}\mathcal{M}_N)
\]
and
\[
p(\alpha, \sigma^2) =p(\alpha)p(\sigma^2) \qquad\mbox{for }\mathcal{M}_N,
\]
where $p(\bbmm{\beta}|\sigma^2, g)$, $p(\alpha)$ and $p(\sigma^2)$
are given by (\ref{prior-beta-1}), (\ref{prior-alpha})
and
(\ref{prior-sigma}), and $p(g)$ when $q<n-1$ is given by
(\ref{prior-g}) with $-1<a<-1/2$ and $b=(n-5)/2-q/2-a$
[$p(g)$ is arbitrary when $q \geq n-1$],
we have a following theorem about the Bayes factor
ratio of the marginal densities under
each of $\mathcal{M}_\gamma$ and $\mathcal{M}_N$.
%
%
\begin{theorem}\label{main-thm-1}
The Bayes factor for comparing
each of $ \mathcal{M}_\gamma$ to $\mathcal{M}_N$ is
%
%
\begin{eqnarray}\label{BC-2}
\BF\bgvn(a,\bbmm{\nu})&=&
\frac{m_\gamma(\bbm{y})}{m_N(\bbm{y})} \nonumber\\[-8pt]\\[-8pt]
&=&
\cases{
\displaystyle \prod_{i=1}^{q} \nu_i^{-1/2}
\frac{B({q/2}+a+1,({n-q-3})/{2}-a)}
{B(a+1,({n-q-3})/{2}-a)}\vspace*{2pt}\cr
\displaystyle\hphantom{\prod_{i=1}^{q} }
{}\times\frac{(1-Q^2)^{-{q}/{2}-a-1}}{(1-R^2)^{({n-q-3})/{2}-a}},
\qquad \mbox{if $q < n-1$}, \vspace*{2pt}\cr
\displaystyle \prod_{i=1}^{n-1} \nu_i^{-1/2}( 1-Q^2)^{-(n-1)/2}, \qquad
\mbox{\hspace*{2.1pt}if $q \geq
n-1$},}\nonumber
\end{eqnarray}
where $ \nu_1 \geq\cdots\geq\nu_r \geq1$, $R^2$ and
$ Q^2$ are given by
%
%
\begin{equation}\label{R2Q2}
R^2
= \sum_{i=1}^r \{\operatorname{cor}(\bbm{u}_i,\bbm{y})\}^2,\qquad
Q^2 = \sum_{i=1}^r (1-\nu_i^{-1})
\{\operatorname{cor}(\bbm{u}_i,\bbm{y})\}^2.
\end{equation}
\end{theorem}

Note that $R^2$ and $Q^2$ are the usual and a modified version
of the $R$-squared statistics and $\operatorname{cor}(\bbm{u}_i,\bbm{y})$
is the correlation of the response $\bbm{y}$
and the $i$th principal component of $\bbm{X}$.
\begin{pf*}{Proof of Theorem \ref{main-thm-1}}
Defining $\bbm{v}=\bbm{y}-\bar{y}\bbm{1}_n$, where $\bar{y}$ is the
mean of $\bbm{y}$,
so that
\[
\| \bbm{y} -\alpha\bbm{1}_n- \bbm{X}\bbmm{\beta} \|^2
= n(-\alpha+\bar{y})^2
+ \|\bbm{v}-\bbm{X} \bbmm{\beta} \|^2,
\]
we obtain
%
%
\begin{equation}\label{marginal-alpha}
\int_{-\infty}^{\infty}
p(\bbm{y}|\alpha,\bbmm{\beta}, \sigma^2)
\,d \alpha= \frac{n^{1/2}}{(2\pi\sigma^2)^{(n-1)/2}}
\exp\biggl(
-\frac{\| \bbm{v}-\bbm{X}\bbmm{\beta}\|^2}{2\sigma^2}\biggr) .
\end{equation}
We make the following orthogonal
transformation when integration with respect to $\bbmm{\beta}$ is considered:
%
%
\begin{equation}\label{btrans}
\bbmm{\beta} \to
\cases{
\bbm{W}'\bbmm{\beta} \equiv\bbmm{\beta}_*, &\quad if $q \leq n-1$, \cr
\pmatrix{
\bbm{W}' \bbmm{\beta} \vspace*{2pt}\cr
\bbm{W}'_\# \bbmm{\beta}}
\equiv
\pmatrix{
\bbmm{\beta}_* \cr
\bbmm{\beta}_\#
},
&\quad if $ q > n-1$,}
\end{equation}
so that
\begin{eqnarray*}
&& \int_{-\infty}^{\infty} \int_{R^q}
p(\bbm{y}|\alpha,\bbmm{\beta}, \sigma^2)p(\bbmm{\beta}|\sigma^2,g)
\,d \alpha \,d\bbmm{\beta} \\[-3pt]
&&\qquad=
\frac{n^{1/2}}{(2\pi\sigma^2)^{(n-1)/2}}
\frac{|\bbmm{\Psi}|^{-1/2}}
{(2\pi\sigma^2 )^{r/2}}
\int_{R^{r}}
\exp\biggl(-\frac{\| \bbm{v}-\bbm{UD}\bbmm{\beta}_* \|^2}{2\sigma^2}
-\frac{\bbmm{\beta}'_*\bbmm{\Psi}^{-1}\bbmm{\beta}_*}{2\sigma
^2}\biggr) \,d\bbmm{\beta}_* \\[-3pt]
&&\qquad\quad{} \times\cases{
1, &\quad if $q \leq n-1$, \vspace*{2pt}\cr
\displaystyle \int_{R^{q-n+1}}p_\#(\bbmm{\beta}_\#)\,d \bbmm{\beta}_\#\
(\mbox{$=$}1), &\quad
if $q > n-1$.}
\end{eqnarray*}
Completing the square
$ \|\bbm{v} - \bbm{UD}\bbmm{\beta}_*\|^2+\bbmm{\beta}'_* \bbmm{\Psi}^{-1}
\bbmm{\beta}_*$
with respect to $\bbmm{\beta}_*$, we have
%
%
\begin{eqnarray}\label{complete-square}
&& \|\bbm{v} - \bbm{UD}\bbmm{\beta}_*\|^2+\bbmm{\beta}'_* \bbmm{\Psi}^{-1}
\bbmm{\beta}_*
\nonumber\\[-2pt]
&&\qquad =\{\bbmm{\beta}_*-(\bbm{D}^2+\bbmm{\Psi}^{-1})^{-1}\bbm{D}'
\bbm{U}'\bbm{v}\}'(\bbm{D}^2+\bbmm{\Psi}^{-1})\nonumber\\[-9pt]\\[-9pt]
&&\qquad\quad{}\times
\{\bbmm{\beta}_*-(\bbm{D}^2+\bbmm{\Psi}^{-1}
)^{-1}\bbm{D}'\bbm{U}'\bbm{v}\} \nonumber\\[-2pt]
&&\qquad\quad{} -\bbm{v}'\bbm{UD}(\bbm{D}^2+\bbmm{\Psi}^{-1})^{-1}\bbm{D}'
\bbm{U}'\bbm{v}+\bbm{v}'\bbm{v},
\nonumber
\end{eqnarray}
where the residual term is rewritten as
\begin{eqnarray*}
&& -\bbm{v}'\bbm{UD}(\bbm{D}^2+\bbmm{\Psi}^{-1})^{-1}\bbm{D}'\bbm{U}'
\bbm{v}+\bbm{v}'\bbm{v} \\[-3pt]
&&\qquad =-\bbm{v}'\Biggl(\sum_{i=1}^r \bbm{u}_i \bbm{u}'_i \frac
{d_i^2}{d_i^2+\psi_i^{-1}}\Biggr)\bbm{v}
+\bbm{v}'\bbm{v} \\[-3pt]
&&\qquad =\frac{g\|\bbm{v}\|^2}{g+1}\Biggl\{ 1- \sum_{i=1}^r \frac
{(\bbm{u}'_i\bbm{v})^2}{\|\bbm{v}\|^2}\Biggr\}
+\frac{\|\bbm{v}\|^2}{1+g} \Biggl\{ 1- \sum_{i=1}^r \biggl(1-\frac
{1}{\nu_i}\biggr)
\frac{(\bbm{u}'_i\bbm{v})^2}{\|\bbm{v}\|^2}\Biggr\}.
\end{eqnarray*}
Hence, by
\[
|\bbmm{\Psi}|=\prod_{i=1}^r \frac{\nu_i+\nu_i g-1}{d_i^2},\qquad
|\bbm{D}^2+\bbmm{\Psi}^{-1}|=\prod_{i=1}^r
\frac{d_i^2\nu_i(1+g)}{\nu_i+\nu_i g-1},
\]
we have
%
%
\begin{eqnarray}\label{marginal-alpha-beta}
&&\int_{-\infty}^{\infty} \int_{R^q}
p(\bbm{y}|\alpha,\bbmm{\beta}, \sigma^2)p(\bbmm{\beta}|g, \sigma^2)
\,d \alpha \,d\bbmm{\beta} \nonumber\\[-9pt]\\[-9pt]
&&\qquad=
\frac{n^{1/2}}{(2\pi\sigma^2)^{(n-1)/2}}
\frac{(1+g)^{-r/2}}{\prod_{i=1}^r \nu_i^{1/2}}
\exp\biggl(
-\frac{\|\bbm{v}\|^2\{g(1-R^2)+1-Q^2\}}{2\sigma^2(g+1)}
\biggr),\hspace*{-25pt}
\nonumber
\end{eqnarray}
where $R^2$ and $Q^2$ are given by (\ref{R2Q2}).\vadjust{\goodbreak}

Next we consider the integration with respect to $\sigma^2$.
By (\ref{marginal-alpha-beta}), we have
%
%
\begin{eqnarray}\label{marginal-alpha-beta-sigma}
&&\int_{-\infty}^{\infty} \int_{R^q} \int_{0}^{\infty}
p(\bbm{y}|\alpha,\bbmm{\beta}, \sigma^2)
p(\bbmm{\beta}|g, \sigma^2)\frac{1}{\sigma^2}
\,d \alpha \,d\bbmm{\beta} \,d\sigma^2 \nonumber\\
&&\qquad = \int_{0}^{\infty}\frac{n^{1/2}}{(2\pi\sigma^2)^{(n-1)/2}}
\frac{(1+g)^{-r/2}}{\prod_{i=1}^r \nu_i^{1/2}}\nonumber\\[-8pt]\\[-8pt]
&&\hphantom{\int_{0}^{\infty}}
\qquad\quad{}\times
\exp\biggl(-\frac{\|\bbm{v}\|^2\{g(1-R^2)+1-Q^2\}}{2\sigma^2(g+1)}
\biggr)
\frac{1}{\sigma^2}\,d\sigma^2
\nonumber\\
&&\qquad = \frac{K(n,\bbm{y})}{\prod_{i=1}^r \nu_i^{1/2}}
(1+g)^{-r/2+(n-1)/2}\{ g(1-R^2)+1-Q^2
\}^{-(n-1)/2},\nonumber
\end{eqnarray}
where
\[
K(n,\bbm{y})=\frac{n^{1/2}\Gamma(\{n-1\}/2)}
{\pi^{(n-1)/2}\|\bbm{y}-\bar{y}\bbm{1}_n\|^{n-1}}.
\]
When $ q \geq n-1$, $R^2=1$ and $r=n-1$ so that
%
%
\begin{eqnarray}\label{marginal-many}
&& \int_{-\infty}^{\infty} \int_{R^q} \int_{0}^{\infty}
p(\bbm{y}|\alpha,\bbmm{\beta}, \sigma^2) p(\bbmm{\beta}|g, \sigma
^2)\frac{1}{\sigma^2}
\,d \alpha \,d\bbmm{\beta} \,d\sigma^2 \nonumber\\[-8pt]\\[-8pt]
&&\qquad = \frac{K(n,\bbm{y})}{\prod_{i=1}^{n-1} \nu_i^{1/2}}
\{ 1-Q^2 \}^{-(n-1)/2},\nonumber
\end{eqnarray}
which does not depend on $g$.
Hence, in this case, $m_\gamma(\bbm{y})$ does not depend on the prior
density of $g$.

When $q < n-1$, we consider the prior (\ref{prior-g}) of $g$
with $-1<a<-1/2$ and $ b =(n-5)/2-q/2-a $,
where $b$ is guaranteed to be strictly greater
than $-1$ for $q < n-1$.
Then we have
%
%
\begin{eqnarray}\label{marginal-few}
m_\gamma(\bbm{y})
&=& \frac{K(n,\bbm{y})}{\prod_{i=1}^q \nu_i^{1/2}B(a+1,b+1)}
\nonumber\\
&&{} \times
\int_0^{\infty} \frac{g^b}{(1+g)^{a+b+2}}
\frac{\{g(1-R^2)+1-Q^2 \}^{-(n-1)/2}}{(1+g)^{q/2-(n-1)/2}}
\,dg \nonumber\\
&=&
\frac{K(n,\bbm{y})(1-Q^2)^{-(n-1)/2}}
{\prod_{i=1}^q \nu_i^{1/2}B(a+1,b+1)}
\int_0^{\infty} g^{b}
\biggl(\frac{1-R^2}{1-Q^2}g+1 \biggr)^{-(n-1)/2} \,dg \\
&=&\frac{K(n,\bbm{y})(1-Q^2)^{-(n-1)/2+b+1}}
{\prod_{i=1}^q \nu_i^{1/2} \{1-R^2\}^{b+1}}
\frac{B(q/2+a+1,b+1)}{B(a+1,b+1)} \nonumber\\
&=&\frac{K(n,\bbm{y})(1-Q^2)^{-q/2-a-1}}
{\prod_{i=1}^q \nu_i^{1/2} \{1-R^2\}^{(n-q-3)/2-a}}
\frac{B(q/2+a+1,(n-q-3)/2-a)}{B(a+1,(n-q-3)/2-a)}.
\hspace*{-20pt} \nonumber
\end{eqnarray}\vfill\eject
In the same way, $m_N(\bbm{y})$ for the null model
is obtained as
%
%
\begin{equation} \label{marginal-null}
m_{N}(\bbm{y}) =K(n,\bbm{y}).
\end{equation}
From (\ref{marginal-many}), (\ref{marginal-few}) and (\ref{marginal-null}),
the theorem follows.\vspace*{2pt}
\end{pf*}
%
%
\begin{remark}
$R^2$ and $Q^2$ given by (\ref{R2Q2}) are the usual and a modified
form of
the $R$-squared measure for multiple regression.
They are here expressed in terms of
$\{\operatorname{cor}(\bbm{u}_1,\bbm{y})\}^2,\ldots, \{\operatorname{cor}
(\bbm{u}_r,\bbm{y})\}^2 $,
the squared correlations of the response $\bbm{y}$ and the principal components
$\bbm{u}_1,\ldots, \bbm{u}_r$ of $\bbm{X}$.
For fixed~$q$ and $\bbmm{\nu}$, the BF criterion is increasing in both
$R^2$ and $Q^2$.
The former is definitely reasonable.
Larger $Q^2$ would also be reasonable when $\nu_1 \geq\cdots\geq\nu_r$
so that $Q^2$ would put more weight on those components of $\bbm{W}'
\bbmm{\beta}$
for which~$d_i$ is larger and are consequently better estimated.
In this sense, $Q^2$ would reward those models which are more stably estimated.

Beyond their influence through $Q^2$, the choice of $\nu_1, \ldots,
\nu_r$ plays a further influential role in $\BF\bgvn$ through the
$\prod_{i=1}^{r} \nu_i^{-1/2}$ terms in (\ref{BC-2}). In Section~%
\ref{secdefaults} below,
a default choice is proposed which, through these terms, rewards stable
estimation.
Note that if $\nu_i=1$ for all $i$ (i.e., the original $g$-prior),
$Q^2$ becomes zero, $\prod_{i=1}^{r} \nu_i^{-1/2} \equiv1$, and $\BF
\bgvn$ becomes a function of just $R^2$ and $q$. In this case, $\BF
\bgvn$ will not distinguish between models for which $q \geq
n-1$.\vspace*{2pt}
\end{remark}
%
%
\begin{remark}
The analytical simplification in (\ref{marginal-few}) is a consequence
of the choice $b = (n-5)/2-q/2-a$, and results in a convenient closed form
for our Bayes factor.
Such a reduction is unavailable for other choices of $b$.
For example, \citet{Liang-etal-2008} use Laplace approximations
to avoid the evaluation of the special functions that arise
in the resulting Bayes factor when $b=0$.
Another attractive feature of the choice $b = (n-5)/2-q/2-a$ will be
discussed in Section~\ref{subsecOn}.\vspace*{3pt}
\end{remark}
%
\subsection{Default choices}\label{secdefaults}
At this point, we are ready to consider default choices for $a$ and
$\bbmm{\nu}$.
For $a$, we recommend
%
%
\begin{equation} \label{our-a}
a =-3/4,
\end{equation}
the median of the range of values $(-1,-1/2)$ for which the marginal
density is well defined for any choices of $q < n-1$.
In Section \ref{sechyper} we will explicitly see the appealing consequence
of this choice on the asymptotic tail behavior of~$p(\bbmm{\beta} | \sigma^2)$.

For $\bbmm{\nu}$, we recommend
%
%
\begin{equation}\label{default-nu}
\bbmm{\nu} =(d^2_1/d^2_r,d^2_2/d^2_r,\ldots, 1)',
\end{equation}
which coupled with (\ref{general-psi}) satisfies (\ref
{weak-des-order}) since $\nu_1 \geq\cdots\geq\nu_q \geq1$ for this
choice. Inserting this $\bbmm{\nu}$ into (\ref{R2Q2}) yields
%
%
\begin{eqnarray}\label{gess-1}
Q^2 &=& R^2
- d_r^2 \sum_{i=1}^r \frac{(\bbm{u}'_i \bbm{v})^2}{d_i^2 \bbm{v}'
\bbm{v}} \nonumber\\
&=& R^2 - d_r^2\bigl\| \bbm{D}^{-1}\bbm{U}'\{\bbm{v}/\|\bbm{v}\|\}
\bigr\|^2 \\
&=&
\cases{
R^2 - d_q^2 \| \hat{\bbmm{\beta}}_{\mathrm{LS}}\|^2, &\quad if $q < n-1$, \vspace*{2pt}\cr
1- d_{n-1}^2 \| \hat{\bbmm{\beta}}_{\mathrm{LS}}^{\mathrm{MP}}\|^2, &\quad if $q \geq n-1$,}
\nonumber
\end{eqnarray}
where, for the standardized response
$\bbm{v}/\|\bbm{v}\|$ for $\bbm{v}=\bbm{y}-\bar{y}\bbm{1}_n$,
$ \hat{\bbmm{\beta}}_{\mathrm{LS}}$ is the usual LS estimator for $q < n-1$,
and $ \hat{\bbmm{\beta}}_{\mathrm{LS}}^{\mathrm{MP}}$ is the LS estimator based on the
Moore--Penrose
inverse matrix.
The third equality in (\ref{gess-1}) follows from the fact that both
$\hat{\bbmm{\beta}}_{\mathrm{LS}}$ and $\hat{\bbmm{\beta}}_{\mathrm{LS}}^{\mathrm{MP}}$
for the response $\bbm{v}/\|\bbm{v}\|$
can be expressed as
\[
\hat{\bbmm{\beta}} = \bbm{W}\bbm{D}^{-1}\bbm{U}'\{\bbm{v}/\|\bbm{v}\|\},
\]
and from the orthogonality of $\bbm{W}$,
\[
\| \hat{\bbmm{\beta}}\|^2= \bigl\| \bbm{D}^{-1}\bbm{U}'\{\bbm{v}/\|
\bbm{v}\|\}\bigr\|^2.
\]
It will also be useful to define
%
%
\begin{equation}
\bar{d} = \Biggl(\prod_{i=1}^r d_i \Biggr)^{1/r},
\end{equation}
the geometric mean of the singular values $d_1,\ldots, d_r$.
Inserting our default choices for $a$ and $\bbmm{\nu}$ into
$\BF\bgvn(a,\bbmm{\nu})$ in (\ref{BC-2}),
and noting that
%
%
\begin{equation}\label{gess-2}
\prod_{i=1}^r \nu_i^{-1/2} = (\bar{d}/d_r)^{-r},
\end{equation}
we obtain our recommended Bayes factor in (\ref{gBF}) which we denote
by $\gbf$
(\mbox{$g$-prior} Bayes factor):
%
%
\begin{eqnarray}\label{gBfcrit}\quad
&& \gbf\bgvn\nonumber\\[-8pt]\\[-8pt]
&&\qquad =
\cases{\displaystyle
\biggl\{\frac{\bar{d}}{d_q}\biggr\}^{-q}
\frac{B({q/2}+{1/4},({n-q})/{2}-{3}/{4})}
{B({1}/{4},({n-q})/{2}-{3}/{4})}\vspace*{2pt}\cr
\displaystyle \qquad{}\times
\frac{(1-R^2+d_{q}^2\|\hat{\bbmm{\beta}}_{\mathrm{LS}}\|^2)^{-
{1}/{4}-{q}/{2}}}
{(1-R^2)^{({n-q})/{2}-{3}/{4}}},
&\quad if $q < n-1$, \cr
\{\bar{d} \times
\| \hat{\bbmm{\beta}}^{\mathrm{MP}}_{\mathrm{LS}}\|\}^{-(n-1)},
&\quad if $q \geq n-1$,}\nonumber
\end{eqnarray}
which is a function of the key quantities $q$, $R^2$,
the LS estimators and the singular values of the design matrix.
%
%
\begin{remark}\label{remark-stability}
$\!\!\!\!$Like traditional selection criteria such as AIC, BIC and~RIC,
the $\gbf$ criterion (\ref{gBfcrit}) rewards models for explained variation
through~$R^2$.
However, $\gbf$ also rewards models for stability of estimation
through smaller values of $\bar{d}/d_q$ and $d_{q} \|\hat{\bbmm{\beta
}}_{\mathrm{LS}}\|$
for $q < n-1$, and through smaller values of the product $\bar{d}/d_{n-1}$
and $d_{n-1} \|\hat{\bbmm{\beta}}_{\mathrm{LS}}^{\mathrm{MP}}\|$ for $q \ge n-1$,
the case where $R^2$ is unavailable.

To see how these various quantities bear on stable estimation, note
first that
%
%
\begin{equation}\label{measure1}
\bar{d}/d_r =\Biggl\{\prod_{i=1}^r(d_i/d_r)\Biggr\}^{1/r},
\end{equation}
which gets smaller as the $d_i/d_r$ ratios get smaller.
Like the well-known condition number $d_1/d_r$, smaller values of (\ref
{measure1})
indicate a more stable design matrix $\bbm{X}_\gamma$.

For $d_{q} \|\hat{\bbmm{\beta}}_{\mathrm{LS}}\|$ and $d_{n-1}\| \hat{
\bbmm{\beta}}_{\mathrm{LS}}^{\mathrm{MP}}\|$,
note that each of these can be expressed as
%
%
\begin{equation} \label{measure2}
d_{r}^2\|\hat{\bbmm{\beta}}\|^2= \sum_{i=1}^r
\biggl(\frac{d_r}{d_i}\biggr)^2
\biggl\{ \frac{(\bbm{u}'_i\bbm{v})}{\|\bbm{u}_i\| \|\bbm{v}\|}
\biggr\}^2
=\sum_{i=1}^r
\biggl(\frac{d_r}{d_i}\biggr)^2
\{\operatorname{cor}(\bbm{u}_i,\bbm{y})\}^2.
\end{equation}
Thus, for a given set of $d_i/d_r$ ratios,
(\ref{measure2}) gets smaller
if the larger correlations $\operatorname{cor}(\bbm{u}_i,\bbm{y})$
correspond to the larger $d_i$.
Again, this is a measure of stability, as the largest principal components
$d_i \bbm{u}_i$ are the ones which are most stably estimated.
\end{remark}
%
%
\begin{remark}\label{remark-dr}
The choice of $\bbmm{\nu}$ in (\ref{default-nu}) will be especially
sensitive to small values of $d_r$ which would lead to large prior
variances in (\ref{prior-beta-1}). Thus, one bad $x_i$ predictor
variable could spoil the model. From an estimation point of view, this
perhaps would be unwise. However, from a model selection point of view,
the effect of a small $d_r$ would have the effect of downweighting the
model, through the stability measures discussed in Remark \ref
{remark-stability}, in favor of models which left out the offending
$x_i$. Thus,
any unstable submodel with at least one such $x_i$, but possibly more,
would be downweighted.
\end{remark}

\section{The effect of the default choices of $a$ and $b$}
\label{sechyper}
In Section \ref{secmarginal+BF}
we proposed the prior form $p(g)$ given by (\ref{prior-g})
with hyperparameters $a$ and $b$, recommending the choices
$a=-3/4$ and $b=(n-q-5)/2-a$ for the case $q < n-1$ where the prior on
$g$ matters.
In the following subsections, we show some appealing consequences of
these choices.

\subsection{\texorpdfstring{The effect of $a$ on the tail behavior of $p(\beta|\sigma^2)$}
{The effect of $a$ on the tail behavior of p(beta|sigma^2)}}
\label{subseca}
Combining $p(\bbmm{\beta}|g, \sigma^2)$ in (\ref{prior-beta-1})
with $p(g)$ in (\ref{prior-g}), the probability density of
$\bbmm{\beta}$ given $\sigma^2$ is given by
%
%
\begin{equation}
p(\bbmm{\beta}|\sigma^2)=
\int_0^\infty
\frac{\phi_q(\bbm{W}'\bbmm{\beta};\bbm{0},\sigma^2\bbmm{\Psi
}_q(g,\bbmm{\nu}))}{B(a+1,b+1)}
\frac{g^b }{(1+g)^{a+b+2}}\,dg.\vadjust{\goodbreak}
\end{equation}
To examine the asymptotic behavior of the density
$p(\bbmm{\beta}|\sigma^2)$ as $\|\bbmm{\beta}\| \to\infty$,
we appeal to the Tauberian theorem for the Laplace transform
[see \citet{geluk-dehaan-1987}],
which tells us that the contribution of the integral (\ref{eqc-nu})
around zero
becomes negligible as $\|\bbmm{\beta}\| \to\infty$.
Thus, we have only to consider the integration between $\nu_1$ and
$\infty$
(the major term).

Since $d_1 \geq\cdots\geq d_q$, and assuming $\nu_1 \geq\cdots\geq
\nu_q$,
we have
%
%
\begin{equation} \label{eqc-nu}
\frac{d_q^2}{(\nu_1+1)g} \leq\frac{d_i^2}{\nu_i+\nu_i g-1} \leq
\frac{d_1^2}{\nu_q g}
\end{equation}
for $g \geq\nu_1$ and any $i$, which implies
\begin{eqnarray*}
&& C\frac{d_q^q}{(\nu_1+1)^{q/2}}
\int_{\nu_1}^\infty
\biggl(\frac{g}{g+1}\biggr)^{a+b+2}\biggl(\frac{1}{g}\biggr)^{q/2+a+2}
\exp\biggl( -\frac{1}{g}
\frac{ d_1^2\|\bbm{W}'\bbmm{\beta}\|^2}
{2\nu_q\sigma^2}
\biggr)\,dg \\
&&\qquad \leq\mbox{the major term of } p(\bbmm{\beta}|\sigma^2) \\
&&\qquad \leq
C\frac{d_1^q}{\nu_q^{q/2}}
\int_{\nu_1}^\infty
\biggl(\frac{g}{g+1}\biggr)^{a+b+2}\biggl(\frac{1}{g}\biggr)^{q/2+a+2}
\exp\biggl( -\frac{1}{g}
\frac{ d_q^2\|\bbm{W}'\bbmm{\beta}\|^2}{2(\nu_1+1)\sigma^2}\biggr)\,dg,
\end{eqnarray*}
where $C=\{B(a+1,b+1)\}^{-1}(2\pi\sigma^2)^{-q/2}$.
Thus, by the Tauberian theorem, there exist $C_1<C_2$ such that
%
%
\begin{equation} \label{tail-behavior-beta}
C_1 <\frac{\|\bbmm{\beta}\|^{q+2a+2}}
{(\sigma^2)^{a+1}}p(\bbmm{\beta}|\sigma^2)
< C_2
\end{equation}
for sufficiently large $\|\bbmm{\beta}\|$.

From (\ref{tail-behavior-beta}), we see
that the asymptotic tail behavior of $p(\bbmm{\beta}|\sigma^2)$ is determined
by $a$ and unaffected by $b$.
Smaller $a$ yields flatter tail behavior, thereby diminishing the prior
influence
of $p(\bbmm{\beta}|\sigma^2)$.
For $a=-1/2$ the asymptotic tail behavior of $p(\bbmm{\beta}|\sigma^2)$,
$\|\bbmm{\beta}\|^{-q-1}$, corresponds to that of
multivariate Cauchy distribution recommended by
\citet{Zellner-Siow-1980}.
In contrast, the asymptotic tail behavior of our choice $a =-3/4$,
$ \|\bbmm{\beta}\|^{-q-1/2}$,
is even flatter than that of the multivariate Cauchy distribution.

\subsection{The effect of $b$ on the implicit $O(n)$ choice of
$g$}\label{subsecOn}

For implementations of the original $g$-prior (\ref{g-prior}),
\citet{Zellner-1986} and others have recommended choices for which
$g = O(n)$.
This prevents the $g$-prior from asymptotically dominating the likelihood
which would occur if $g$ was unchanged as $n$ increased.
The recommendation of choosing $g = O(n)$ also applies to the choice of
a fixed $g$
for the generalized $g$-prior (\ref{prior-beta-1}) where
\[
\operatorname{tr}\{\operatorname{Var}(\bbmm{\beta}|g,\sigma^2)\}= \sigma^2
\sum_{i=1}^q \frac{\nu_i+\nu_i g-1}{d_i^2}.
\]
Since $d_i^2=O(n)$ for $1 \leq i \leq q$ by Lemma \ref{lemmaprel-1},
$ \operatorname{tr}\{\operatorname{Var}(\bbmm{\beta}|g,\sigma^2)\}=g O(n^{-1})$
if~$ \nu_i$ is bounded.
Therefore, the choice $g=O(n)$ will also prevent the\vadjust{\goodbreak}
generalized $g$-prior from asymptotically dominating the likelihood,
and stabilize it in the sense that
$\operatorname{tr}\{\operatorname{Var}(\bbmm{\beta}|g,\sigma^2)\}= O(1)$ when
$g=O(n)$.

For our fully Bayes case, where $g$ is treated as a random variable,
our choice of $b$, in addition to yielding a closed form for the
marginal density
in~(\ref{marginal-few}), also yields an implicit $O(n)$ choice of $g$,
in the sense that
\begin{eqnarray*}
[\mbox{mode of }g]&=& \frac{b}{a+2}= \frac{2(n-q)-7}{5}, \\
\frac{1}{E[g^{-1}]} &=& \frac{b}{a+1}=2(n-q)-7
\end{eqnarray*}
for our recommended choices $a=-3/4$ and $b=(n-q-5)/2-a$.
(Note that~$E[g]$ does not exist under the choice $a=-3/4$.)

\section{Shrinkage estimation conditionally on a model} \label{secafter}
In this section we consider estimation conditionally on a model
$\mathcal{M}_\gamma$.
Because $\bbmm{\beta}$ is not identifiable when $q > n-1$, and hence
not estimable,
we instead focus on estimation of~$\bbm{X}\bbmm{\beta}$, 
which is always estimable.
For this purpose, we consider estimation of~$\bbm{X}\bbmm{\beta}$ under
scaled quadratic loss
$(\bbmm{\delta} - \bbm{X}\bbmm{\beta})'\bbm{Q}(\bbmm{\delta}-\bbm
{X}\bbmm{\beta
})/\sigma^2$
for positive-definite~$\bbm{Q}$.
The Bayes estimator under this loss for any $\bbm{Q}$ is of the form
%
%
\begin{equation} \label{xb}
\bbm{X} \hat{\bbmm{\beta}}_B = \bbm{X} E[\sigma^{-2}\bbmm{\beta}|
\bbm{y}]/E[\sigma^{-2}|\bbm{y}].
\end{equation}
From calculations similar to those in Section \ref{secmarginal+BF},
under our priors given in Section~\ref{secpriors},
a simple closed form can be obtained for this estimator as follows.
In contrast, such a simple closed form is not available for the usual
Bayes estimator, $\bbm{X}E[\bbmm{\beta}_\gamma|\bbm{y}]$, the
posterior mean under $(\bbmm{\delta} - \bbm{X}\bbmm{\beta})'\bbm{Q}(
\bbmm{\delta}-\bbm{X}\bbmm{\beta})$ which does not scale for the
variance~$\sigma^2$.
%
%
\begin{theorem}\label{thmestimator}
The Bayes estimator under scaled quadratic loss is given by
%
%
\begin{equation} \label{reasonable-1}
\bbm{X} \hat{\bbmm{\beta}}_B
= \sum_{i=1}^r \bigl(1-H(\bbm{y})/\nu_i \bigr)(\bbm{u}'_i \bbm
{y})\bbm{u}_i,
\end{equation}
where
\[
H(\bbm{y})=
\cases{
\biggl( 1+\dfrac{1-Q^2}{1-R^2} \dfrac{(n-q-3)/2-a}{q/2+a+1}
\biggr)^{-1}, &\quad $q<n-1$, \vspace*{2pt}\cr
\{1+E[g]\}^{-1}, &\quad $q\geq n-1$.}
\]
\end{theorem}
\begin{pf}
See the \hyperref[app]{Appendix}.
\end{pf}

Thus, when $q \geq n-1$, we must specify the mean of prior density of
$g$, although
no such specification was needed for model selection.
A reasonable specification may be $E[g]=d^2_{n-1}/d^2_1$, a function
of\vadjust{\goodbreak}
the condition number $ d_1/d_{n-1}$ of the linear equation.
For extremely large values of $d_1/d_{n-1}$,
the coefficients of the first and the last terms in (\ref{reasonable-1})
become nearly $1$ and $0$, respectively.
See \citet{Casella-1985} and \citet{Maru-Straw-2005} for further
discussion of
the condition number.

Thus, for our recommended choices of hyperparameters
$a=-3/4$ and $\nu_i= d_i^2/d_r^2$ for $1 \leq i \leq r$,
our recommended estimator of $\bbm{X}\bbmm{\beta}$ for a given model
$\mathcal{M}_\gamma$ is
%
%
\begin{equation}\label{default-est}
\bbm{X} \hat{\bbmm{\beta}}_B =\sum_{i=1}^{r} \bigl(1-\{
d_{r}^2/d_i^2\}H(\bbm{y})\bigr)
(\bbm{u}'_i \bbm{y})\bbm{u}_i,
\end{equation}
where
%
%
\begin{equation}\quad
H(\bbm{y}) =
\cases{
\biggl( 1+ \dfrac{1-R^2+ d_q^2 \|\hat{\bbmm{\beta}}_{\mathrm{LS}}\|^2}
{1-R^2} \dfrac{n/2-q/2-3/4}{q/2+1/4}
\biggr)^{-1}, \vspace*{2pt}\cr
\hspace*{80pt}\qquad \mbox{if $q < n-1$}, \vspace*{2pt}\cr
(1+d_{n-1}^2/d_1^2)^{-1}, \qquad \mbox{if $q \geq n-1$}.}
\end{equation}
%
%
\begin{remark}
As mentioned in Remark \ref{remark-dr}, a small value of $d_r$ could
be problematic for estimation. This is reflected in (\ref
{default-est}) where a small $d_r$ would diminish overall shrinkage.
However, the probability of such a model would be severely downweighted
in the model selection context, and so this diminished shrinkage would
be of little consequence.
\end{remark}

\section{Model selection consistency}\label{secconsistency}
In this section we consider the model selection consistency in the case
where $p$ is fixed and $n$ approaches infinity.
Posterior consistency for model choice means
\[
\plim_{n \to\infty}
\operatorname{Pr}(\mathcal{M}_T |y)=1 \qquad\mbox{when }\mathcal{M}_T
\mbox{ is the true model},
\]
where plim denotes convergence in probability under the true model
$\mathcal{M}_T$, namely,
$\bbm{y}=\alpha_T \bbm{1}_n+ \bbm{X}_T\bbmm{\beta}_T+\bbmm{\varepsilon}$,
where
$ \bbm{X}_T$ is the $n \times q_T$ true design matrix and
$\bbmm{\beta}_T$ is the true ($q_T\times1$) coefficient vector
and $\bbmm{\varepsilon}_n \sim N_n(\bbm{0},\sigma^2\bbm{I}_n)$.

Let us show that our general criterion, $\BF\bgvn(a,\bbmm{\nu})$
given by
(\ref{BC-2}) with bounded~$\nu_1$, is model selection consistent.
This is clearly equivalent to
%
%
\begin{equation} \label{equiv-consistency}
\plim_{n \to\infty}
\frac{\BF\bgvn(a,\bbmm{\nu})}
{\BF_{T:N}(a,\bbmm{\nu})}
=0 \qquad\forall\mathcal{M}_{\gamma} \neq\mathcal{M}_T .
\end{equation}
Recall that we have already assumed that
$ \bbm{x}'_i\bbm{1}_n=0 $ and $\bbm{x}'_i\bbm{x}_i/n=1$ for any \mbox{$1 \leq
i \leq p$}.
To obtain model selection consistency, we also assume the following:

{\renewcommand\thelonglist{(A\arabic{longlist})}
\renewcommand\labellonglist{\thelonglist}
\begin{longlist}
\item\label{AS1}
The correlation between $x_i$ and $x_j$,
$\bbm{x}'_i\bbm{x}_j/n$, has a limit as $n \to\infty$.
\item\label{AS2}
The limit of the correlation matrix of $x_1,\ldots,x_p$,
$\lim_{n \to\infty}\bbm{X}'_F\bbm{X}_F/n$,
is positive definite.\vadjust{\goodbreak}
\end{longlist}}

\noindent Assumption \ref{AS1} is the standard assumption which also appears in
\citet{Knight-Fu-2000}
and \citet{Zou-2006}. Assumption \ref{AS2} is natural because the columns of
$\bbm{X}_F$
are assumed to be linearly independent.

Our main consistency theorem is as follows.
Note that our recommended choice $\nu_1=d_1^2/d_q^2$ is bounded
by Lemma \ref{lemmaprel-1} in the \hyperref[app]{Appendix}.
%
%
\begin{theorem} \label{thmconsistency}
Under assumptions \ref{AS1} and \ref{AS2}, if $\nu_1$ is bounded, then
$\BF\bgvn(a,\bbmm{\nu})$ is consistent for
model selection.
\end{theorem}

\section{Simulated performance evaluations}\label{secsim}
$\!\!$In this section we report on a~number of simulated performance
comparisons between
our recommended Bayes factor $\gbf\bgvn$
and the following selection criteria:
\begin{eqnarray*}
\mathrm{ZE} &=&
(1-R^2)^{-(n-q)/2+3/4}
\frac{B(q/2+1/4,(n-q)/2-3/4)}{B(1/4,(n-q)/2-3/4)}, \\
\mathrm{EB} &=& \max_g m_\gamma(\bbm{y}|g,\hat\sigma^2),\\
\mathrm{AIC} &=& -2 \times\mbox{maximum log likelihood} +2(q+2), \\
\mathrm{AICc} &=& -2 \times\mbox{maximum log likelihood} +2(q+2)\frac
{n}{n-q-3}, \\
\mathrm{BIC} &=& -2 \times\mbox{maximum log likelihood} +q \log n.
\end{eqnarray*}
Here, ZE is the special case of $\BF\bgvn$ with
$a=-3/4$ and $\nu_1=\cdots=\nu_q=1$ (corresponding to Zellner's $g$-prior).
Note that comparisons of $\gbf$ with ZE
should reveal the effect of our choice of descending $\nu$.
EB is the empirical Bayes criterion of \citet{George-Foster-2000}
in (\ref{marginal-known-1}), also based on the original $g$-prior,
with $\hat\sigma^2 = \mathrm{RSS}_\gamma/(n-q_\gamma-1)$ plugged in.
Finally, AICc is the well-known correction of AIC proposed by
\citet{Hurvich-Tsai-1989}.

For these comparisons, we consider data generated by submodels
(\ref{submodel-gamma}) of~(\ref{full-model}) with $p = 16$ potential
predictors
for two different choices of the underlying design matrix $\bbm{X}_F$.
For the first choice, which we refer to as the correlated case,
each row of the 16 predictors are generated as
$x_1,\ldots, x_{13} \sim N(0,1)$, and $x_{14}, x_{15}, x_{16} \sim U(-1,1)$
(the uniform distribution) with the following pairwise correlations:
%
%
\begin{equation}
\overbrace{x_1, x_2}^{\operatorname{cor}=0.9},
\underbrace{x_3, x_4}_{\operatorname{cor}=-0.7},
\overbrace{x_5, x_6}^{\operatorname{cor}=0.5},
\underbrace{x_7, x_8}_{\operatorname{cor}=-0.3},
\overbrace{x_9, x_{10}}^{\operatorname{cor}=0.1}
\end{equation}
and independently otherwise.
For the second choice, which we refer to as the simple case, each row
of the 16 predictors are generated as $x_1,\ldots, x_{16}$ i.i.d. $\sim N(0,1)$.

%
%
\begin{table}
\caption{Rank of the true model}\label{table-selected}
\begin{tabular*}{\tablewidth}{@{\extracolsep{\fill}}lcccccccc@{}}
\hline
\hspace*{25pt}$\bolds{q_T}$\textbf{:}\hspace*{-25pt} & \multicolumn{2}{c}{\textbf{16}} & \multicolumn{2}{c}{\textbf{12}} &
\multicolumn{2}{c}{\textbf{8}} & \multicolumn{2}{c@{}}{\textbf{4}} \\[-4pt]
& \multicolumn{2}{c}{\hrulefill} & \multicolumn{2}{c}{\hrulefill} &
\multicolumn{2}{c}{\hrulefill} & \multicolumn{2}{c@{}}{\hrulefill} \\
\hspace*{15pt}\textbf{Rank:}\hspace*{-15pt} & \textbf{1st} & \textbf{1st--3rd}
& \textbf{1st} & \textbf{1st--3rd} & \textbf{1st}
& \textbf{1st--3rd} & \textbf{1st} &
\textbf{1st--3rd} \\
\hline
\multicolumn{9}{@{}c@{}}{Correlated case} \\
[4pt]
$\gbf$ & 0.71 & 0.91 & 0.73 & 0.94 & 0.69 & 0.87 & 0.66 & 0.86 \\
ZE & 0.40 & 0.70 & 0.63 & 0.89 & 0.68 & 0.89 & 0.67 & 0.87 \\
EB & 0.41 & 0.71 & 0.63 & 0.90 & 0.67 & 0.88 & 0.66 & 0.85 \\
AIC & 0.95 & 0.99 & 0.23 & 0.38 & 0.09 & 0.17 & 0.05 & 0.08 \\
AICc& 0.25 & 0.45 & 0.67 & 0.90 & 0.52 & 0.75 & 0.25 & 0.44 \\
BIC & 0.88 & 0.98 & 0.41 & 0.65 & 0.31 & 0.43 & 0.23 & 0.42 \\
[4pt]
\multicolumn{9}{@{}c@{}}{Simple case} \\
[4pt]
$\gbf$ & 0.98 & 0.99 & 0.83 & 0.97 & 0.75 & 0.93 & 0.67 & 0.85 \\
ZE & 0.94 & 0.98 & 0.87 & 0.97 & 0.78 & 0.95 & 0.69 & 0.88 \\
EB & 0.95 & 0.98 & 0.87 & 0.98 & 0.76 & 0.95 & 0.65 & 0.87 \\
AIC & 1.00 & 1.00 & 0.22 & 0.37 & 0.08 & 0.13 & 0.05 & 0.08 \\
AICc & 0.82 & 0.87 & 0.85 & 0.97 & 0.55 & 0.80 & 0.24 & 0.46 \\
BIC & 0.99 & 1.00 & 0.41 & 0.65 & 0.27 & 0.46 & 0.22 & 0.39 \\
\hline
\end{tabular*}
\end{table}

For our first set of comparisons, we set $n=30$ (larger than $p=16$)
and considered 4 submodels where the true predictors are:
\begin{itemize}
\item$x_1, x_2, x_3, x_4, x_5, x_6, x_7, x_8, x_9, x_{10},
x_{11}, x_{12}, x_{13}, x_{14}, x_{15}, x_{16}$ ($q_{T}=16$),\vadjust{\goodbreak}
\item$x_1, x_2, x_3, x_4, x_5, x_6, x_7, x_8, x_{9}, x_{10}, x_{11},
x_{14}$ ($q_{T}=12$),
\item$x_1, x_2, x_5, x_6, x_9, x_{10}, x_{11} , x_{14}$ ($q_{T}=8$),
\item$x_1, x_2, x_5, x_6$ ($q_{T}=4$)
\end{itemize}
(where $q_T$ denotes the number of true predictors) and the true model
is given by
%
%
\begin{equation}\label{sim-model}
Y=1+ 2 \sum_{i \in\{\mathrm{true}\}}x_i+ \{\mbox{normal error term }
N(0,1)\}.
\end{equation}
In both cases, after generating pseudo random $x_1,\ldots,x_{16}$,
we centered and scaled them as noted in Section \ref{secintro}.
%

%
%
\begin{table}
\tabcolsep=0pt
\caption{Prediction error comparisons}\label{table-prediction}
\begin{tabular*}{\tablewidth}{@{\extracolsep{4in minus 4in}}lcccccccc@{}}
\hline
& \multicolumn{2}{c}{\textbf{16}} & \multicolumn{2}{c}{\textbf{12}} &
\multicolumn{2}{c}{\textbf{8}} & \multicolumn{2}{c@{}}{\textbf{4}} \\[-4pt]
& \multicolumn{2}{c}{\hrulefill} & \multicolumn{2}{c}{\hrulefill} &
\multicolumn{2}{c}{\hrulefill} & \multicolumn{2}{c@{}}{\hrulefill} \\
& \textbf{Mean} & \textbf{(LQ, UQ)} & \textbf{Mean}
& \textbf{(LQ, UQ)} & \textbf{Mean} & \textbf{(LQ, UQ)} & \textbf{Mean}
& \textbf{(LQ, UQ)} \\
\hline
\multicolumn{9}{@{}c@{}}{Correlated case} \\
[4pt]
Oracle & 0.57 & (0.43, 0.68) & 0.43 & (0.31, 0.53) & 0.30 & (0.20,
0.38) & 0.17 & (0.09, 0.22) \\
[4pt]
$\gbf$ & 0.70 & (0.44, 0.78) & 0.52 & (0.32, 0.61) & 0.37 & (0.22,
0.47) & 0.26 & (0.11, 0.35) \\
ZE & 1.02 & (0.53, 1.20) & 0.59 & (0.35, 0.71) & 0.41 & (0.23, 0.53) &
0.27 & (0.11, 0.37) \\
EB & 1.00 & (0.52, 1.16) & 0.58 & (0.35, 0.70) & 0.41 & (0.23, 0.53) &
0.27 & (0.11, 0.37) \\
AIC & 0.56 & (0.42, 0.67) & 0.54 & (0.40, 0.65) & 0.51 & (0.37, 0.62) &
0.48 & (0.33, 0.59) \\
AICc & 1.29 & (0.65, 1.65) & 0.56 & (0.34, 0.68) & 0.42 & (0.25, 0.52)
& 0.36 & (0.22, 0.47) \\
BIC & 0.58 & (0.42, 0.69) & 0.53 & (0.38, 0.64) & 0.46 & (0.31, 0.58) &
0.39 & (0.23, 0.51) \\
[4pt]
\multicolumn{9}{@{}c@{}}{Simple case} \\
[4pt]
Oracle & 0.57 & (0.43, 0.68) & 0.43 & (0.31, 0.53) & 0.30 & (0.20,
0.38) & 0.17 & (0.09, 0.22) \\
[4pt]
$\gbf$ & 0.57 & (0.41, 0.67) & 0.45 & (0.33, 0.56) & 0.35 & (0.21,
0.45) & 0.25 & (0.12, 0.33) \\
ZE & 0.66 & (0.42, 0.70) & 0.45 & (0.32, 0.56) & 0.34 & (0.21, 0.44) &
0.24 & (0.12, 0.32) \\
EB & 0.65 & (0.42, 0.69) & 0.45 & (0.32, 0.56) & 0.35 & (0.21, 0.45) &
0.25 & (0.12, 0.34) \\
AIC & 0.56 & (0.42, 0.67) & 0.54 & (0.39, 0.65) & 0.51 & (0.37, 0.63) &
0.48 & (0.32, 0.60) \\
AICc & 0.98 & (0.45, 0.83) & 0.46 & (0.33, 0.55) & 0.39 & (0.25, 0.50)
& 0.35 & (0.20, 0.47) \\
BIC & 0.56 & (0.42, 0.67) & 0.52 & (0.37, 0.64) & 0.45 & (0.30, 0.57) &
0.38 & (0.21, 0.50) \\
\hline
\end{tabular*}
\end{table}

\begin{remark}
With simulations of performance in Bayesian model selection, the
answers primarily depend on the assumed prior. Here we have chosen all
the $\beta_i = 2$, an extreme form of the assumption of exchangeability.
\end{remark}

Table \ref{table-selected} compares the criteria
by how often the true model was selected as best, or in the top 3,
among the $2^{16} $ candidate models across the $N=500$ replications.
We note the following:
\begin{itemize}
\item In the correlated cases, EB, ZE and $\gbf$ were very similar for
$q_T=4,8$, but $g$BF was much better for $q=12, 16$.
\item In the simple cases, $\gbf$, ZE and EB were very similar,
suggesting no effect of
our extension of Zellner's $g$-prior with descending $\nu$.
\item In both the correlated and simple cases, AIC and BIC were poor
for all cases except $q_T=16$.
\item In both the correlated and simple cases, AICc was poor for
$q_T=16 \mbox{ and }4$ but good for $q_T=8, 12$.
\end{itemize}
Overall, Table \ref{table-selected} suggests that $\gbf$ is stable
and good for most cases, and that our generalization of Zellner's
$g$-prior is
effective in the correlated case.

On data from the same setup with $n=30$ and $N=500$,
Table \ref{table-prediction} compares the models selected
by each criterion based on their (in-sample) predictive error
\[
\frac{(\hat{y}_*-\alpha_T1_n- X_T\bbmm{\beta}_T)'
(\hat{y}_*-\alpha_T1_n- X_T\bbmm{\beta}_T)}{n\sigma^2},
\]
where $ X_T$, $\alpha_T$ and $\bbmm{\beta}_T$ are the true $n \times
q_T$ design
matrix, the true intercept and the true coefficients.
The prediction $\hat{y}_{*}$ for each selected model is given
by $\bar{y}1_n+X_{\gamma*}\hat{\bbmm{\beta}}_{\gamma*}$,
where $ X_{\gamma*}$ is the selected design matrix,
$\hat{\bbmm{\beta}}_{\gamma*}$ is the Bayes
estimator for $\gbf$, ZE and EB, and is the least squares estimator
for AIC, BIC and AICc.
To aid in gauging these comparisons, we also included the ``oracle''
prediction error,
namely, that based on the least squares estimate under the true model.

The summary statistics reported in Table \ref{table-prediction}
are the mean predictive error, and
the lower quantile (LQ) and upper quantile (UQ) of the predictive
errors.
In terms of predictive performance, the comparisons are similar to those
in Table \ref{table-selected}.
Overall, we see that $\gbf$ works well in this setting.

For our final evaluations, we use data again simulated from the simple
form (\ref{sim-model}), but now with $x_1, x_2,\ldots, x_{12}$,
$x_{14}, x_{15}$ as the true predictors ($q_T = 14$) and a small sample
size $n=12$ (smaller than $p = 16$). Since $p>q_T>n$, the true model is
not identifiable here. Furthermore, AIC, BIC, AICc, ZE and EB cannot
even be computed (because $p>n$) and so we confine our evaluations to
$\gbf$.

For this very difficult selection situation, $\gbf$ did not rank the
complete true model of dimension $q_T = 14$ as best even once across
the $N=500$ iterations. In fact, as shown by the frequency of model
sizes selected as best by gBF in Table~\ref{selected-number}, the top
selected model was always of dimension less than $n = 12$, the
dimension required for identifiability. However, if one instead
considers the overall $\gbf$ rankings across all possible models, a
different picture emerges.

%
%
\begin{table}
\caption{Model size frequencies in the many predictors case}
\label{selected-number}
\begin{tabular*}{\tablewidth}{@{\extracolsep{\fill}}lccccccc@{}}
\hline
& \textbf{0--6} & \textbf{7} & \textbf{8} & \textbf{9} & \textbf{10}
& \textbf{11} & \textbf{12--16}\\
\hline
Correlated & 0.10 & 0.11 & 0.22 & 0.34 & 0.16 & 0.07 & 0.00 \\
Simple & 0.11 & 0.15 & 0.21 & 0.33 & 0.14 & 0.06 & 0.00 \\
\hline
\end{tabular*}
\vspace*{-6pt}
\end{table}

As can be seen in Table \ref{relative-rank}, which summarizes the
relative rank of the true model
($\operatorname{rank}/2^{16}$) over the $N=500$ iterations
(smaller is better), $\gbf$ often ranked the true model relatively high.
Indeed, the mean relative gBF rank of the true model was $0.035$ in the
correlated case
and $0.039$ in the simple structure case. Both of these mean ranks were
the highest mean ranks
achieved by any of the $2^{16} = 65\mbox{,}536$ candidate models! The true
model ranks were evidently more stable than the other model ranks which
varied more from iteration to iteration. Rather than select a single
top ranked model in this context, it would seem to be better to use
$\gbf$ to restrict interest to a~promising subset.

%
%
\begin{table}
\tablewidth=304pt
\caption{The relative rank of the true model}\label{relative-rank}
\begin{tabular*}{\tablewidth}{@{\extracolsep{\fill}}lcccccc@{}}
\hline
& \textbf{Min} & \textbf{LQ} & \textbf{Median} & \textbf{Mean}
& \textbf{UQ} & \textbf{Max} \\
\hline
Correlated & 0.001 & 0.012 & 0.023 & 0.035 & 0.042 & 0.518 \\
Simple & 0.001 & 0.013 & 0.023 & 0.039 & 0.043 & 0.555 \\
\hline
\end{tabular*}
\end{table}

Further, it should be noted that $\gbf$ performed best among the
larger unidentified models as shown by Table \ref{14}, which reports
the frequencies with which the true model was ranked highly
among the $(16 \times15)/2=120$ candidate models with exactly 14
predictors. To our knowledge, we know of no other analytical selection
criterion for choosing between models with $R^2=1$, which is the case here.

%
%
\begin{table}
\tablewidth=304pt
\caption{Frequency that the true model was ranked highly among models
with $14$ predictors}\label{14}
\begin{tabular*}{\tablewidth}{@{\extracolsep{\fill}}lccc@{}}
\hline
& \textbf{1st} & \textbf{1st--2nd} & \textbf{1st--3rd} \\
\hline
Correlated & 0.14 & 0.22 & 0.26 \\
Simple & 0.13 & 0.20 & 0.26 \\
\hline
\end{tabular*}
\end{table}

Finally, we call attention to Table \ref{each-predictors} which
reports the observed $\gbf$ predictor selection frequencies across the
%
%
\begin{table}
\caption{Predictor frequencies in the many predictors case}
\label{each-predictors}
\begin{tabular*}{\tablewidth}{@{\extracolsep{\fill}}lcccccc@{}}
\hline
& $\bolds{x_1}$ \textbf{(T)}&$\bolds{x_2}$ \textbf{(T)}
& $\bolds{x_3}$ \textbf{(T)} & $\bolds{x_4}$ \textbf{(T)}&
$\bolds{x_5}$ \textbf{(T)} & $\bolds{x_6}$ \textbf{(T)} \\
\hline
Correlated & 0.65 & 0.63 & 0.44 & 0.46 & 0.62 & 0.60 \\
Simple & 0.54 & 0.54 & 0.54 & 0.54 & 0.54 & 0.57 \\
\hline
& $\bolds{x_7}$ \textbf{(T)} & $\bolds{x_8}$ \textbf{(T)}
& $\bolds{x_9}$ \textbf{(T)}& $\bolds{x_{10}}$ \textbf{(T)}
& $\bolds{x_{11}}$ \textbf{(T)} &
$\bolds{x_{12}}$ \textbf{(T)} \\
\hline
Correlated & 0.56 & 0.56 & 0.59 & 0.58 & 0.58 & 0.60 \\
Simple & 0.55 & 0.55 & 0.54 & 0.56 & 0.52 & 0.50 \\
\hline
& $\bolds{x_{13}}$ \textbf{(F)} & $\bolds{x_{14}}$ \textbf{(T)}
& $\bolds{x_{15}}$ \textbf{(T)} & $\bolds{x_{16}}$ \textbf{(F)} & & \\
\hline
Correlated & 0.40 & 0.43 & 0.45 & 0.40 & & \\
Simple & 0.34 & 0.55 & 0.57 & 0.39 & & \\
\hline
\end{tabular*}
\end{table}
top ranked $\gbf$ models over the $N =500$ iterations.
These frequencies show that the top $\gbf$ models tended to at least
be partially correct in the sense that,
for the most part, the true individual predictors [designated by (T)]
were selected more often than not.\looseness=1
%

\begin{remark}
The only variables that were under-selected by $\gbf$ in Table~\ref
{each-predictors}
were $ (x_3, x_4)$ and $(x_{14}, x_{15})$ in the correlated case.
Although $x_3$ and $x_4$ are true predictors, their under-selection may
be explained
by the high negative correlation between them.
Interestingly, the under-selection of $x_{14}$ and $x_{15}$ is not explained
by correlation (as they are independent in both the correlated and
simple cases).
Rather, since all predictors have been standardized, it suggests that
in this setting, selection of $U(-1,1)$ predictors may be more
difficult than $N(0,1)$ predictors (they are uniform in the correlated
case and normal in the simple case).
\end{remark}

%
%
\begin{appendix}\label{app}

\section{\texorpdfstring{Proof of Theorem \lowercase{\protect\ref{thmestimator}}}{Proof of Theorem 5.1}}\label{appA}
We proceed by finding a simple closed form for $\hat{\bbmm{\beta}}_B$
in (\ref{xb}).
Making use of the transformation (\ref{btrans}), and
by the calculation in (\ref{complete-square}),
$E[\bbmm{\beta}_{\#}|\bbm{y}]=E[\bbmm{\beta}_{\#}]$ (say, $\bbmm{\mu
}_{\#}$) and
\begin{eqnarray*}
\bbm{W} \frac{E[\sigma^{-2}\bbmm{\beta}_*|\bbm{y}]}{E[\sigma
^{-2}|\bbm{y}]} &=&
\frac{1}{E[\sigma^{-2}|\bbm{y}]}
E\Biggl[\sigma^{-2}\sum_{i=1}^r\frac{\bbm{u}'_i\bbm{y}}{d_i}
\biggl\{1-\frac{1}{\nu_i(1+g)}\biggr\}\bbm{w}_i\Big|\bbm{y}\Biggr] \\
&=& \sum_{i=1}^r \frac{\bbm{u}'_i\bbm{y}}{d_i} \biggl\{1-\frac{H(\bbm
{y})}{\nu_i}
\biggr\}\bbm{w}_i,
\end{eqnarray*}
where
%
%
\begin{equation}
H(\bbm{y})=\frac{E[\sigma^{-2}(1+g)^{-1}|\bbm{y}]}{E[\sigma^{-2}|\bbm{y}]}.
\end{equation}
Thus,
%
%
\begin{equation}\label{reasonable}
\hat{\bbmm{\beta}}_B = \sum_{i=1}^{r}
\frac{\bbm{u}'_i\bbm{y}}{d_i} \biggl(1-\frac{H(\bbm{y})}{\nu_i}
\biggr)\bbm{w}_i +
\cases{
\bbm{0}, &\quad if $q \leq n-1$, \cr
\bbm{W}_\#\bbmm{\mu}_{\#}, &\quad if $q > n-1$.}
\end{equation}
Since $\bbmm{\beta}$ is not identifiable when $ q \geq n-1$, it is not
surprising that $\hat{\bbmm{\beta}}_B$ is incompletely defined
due to the arbitrariness of $ \bbm{W}_\#\bbmm{\mu}_{\#}$.
However, because $\bbm{X}\bbm{W}_\# = \bbm{0}$, this arbitrariness is
not an issue for the estimation of $\bbm{X}\bbmm{\beta}$, for which we obtain
%
%
\begin{equation}
\bbm{X} \hat{\bbmm{\beta}}_B = \sum_{i=1}^r (\bbm{u}'_i \bbm{y})\bbm{u}_i
\biggl(1-\frac{H(\bbm{y})}{\nu_i} \biggr).
\end{equation}

It now only remains to obtain a closed form for $H(\bbm{y})$.
As in
(\ref{marginal-alpha}), (\ref{marginal-alpha-beta})
and (\ref{marginal-alpha-beta-sigma}) in Section \ref{secmarginal+BF},
%
%
\begin{eqnarray} \label{marginal-alpha-beta-sigma-sigma}
&&\int_{-\infty}^{\infty} \int_{R^q}
\int_{0}^{\infty}\frac{1}{\sigma^2}
p(\bbm{y}|\alpha,\bbmm{\beta}, \sigma^2)p(\bbmm{\beta}|g, \sigma^2)
\frac{1}{\sigma^2}
\,d \alpha \,d\bbmm{\beta} \,d\sigma^2 \nonumber\\
&&\qquad= \int_{0}^{\infty}
\{\sigma^2\}^{-(n+1)/2} \frac{n^{1/2}}{(2\pi)^{(n-1)/2}}
\frac{(1+g)^{-r/2}}{\prod_{i=1}^r \nu_i^{1/2}} \nonumber\\
&&\hphantom{\int_{0}^{\infty}}
\qquad\quad{} \times\exp\biggl(
-\frac{\|\bbm{v}\|^2\{g(1-R^2)+1-Q^2\}}{2\sigma^2(g+1)} \biggr)
\frac{1}{\sigma^2}\,d\sigma^2 \\
&&\qquad = \frac{2n^{1/2}\Gamma(\{n+1\}/2)}{\pi^{(n-1)/2}}
\frac{\|\bbm{v}\|^{-n-1}}{\prod_{i=1}^r \nu_i^{1/2}}
(1+g)^{-r/2+(n+1)/2} \nonumber\\
&&\qquad\quad{} \times\{ g(1-R^2)+1-Q^2
\}^{-(n+1)/2},\nonumber
\end{eqnarray}
which differs slightly from (\ref{marginal-alpha-beta-sigma}) because
of the
extra $1/\sigma^2$ term in the first expression.
Letting
%
%
\begin{equation}
L(\bbm{y}|g)=(1+g)^{-r/2+(n+1)/2} \{ g(1-R^2)+1-Q^2
\}^{-(n+1)/2},
\end{equation}
we have
\begin{eqnarray*}
H(\bbm{y})
&=& \frac{\int_0^\infty(1+g)^{-1} L(\bbm{y}|g) p(g)\,dg}
{\int_0^\infty L(\bbm{y}|g) p(g)\,dg} \\
&=&\frac{\int_0^\infty(1+g)^{-r/2+(n-1)/2}
\{g(1-R^2)+1-Q^2\}^{-(n+1)/2}p(g)\,dg}
{\int_0^\infty(1+g)^{-r/2+(n+1)/2}
\{g(1-R^2)+1-Q^2\}^{-(n+1)/2}p(g)\,dg}.
\end{eqnarray*}

When $ q < n-1$, under the prior (\ref{prior-g})
used in Section \ref{secmarginal+BF}, namely,
\[
p(g)=\frac{g^b(1+g)^{-a-b-2}}{B(a+1,b+1)}=
\frac{g^b(1+g)^{-(n-r-1)/2}}{B(a+1,b+1)},
\]
where $b=(n-5)/2-r/2-a$, we have
\begin{eqnarray*}
H(\bbm{y})&=&\frac{\int_0^\infty g^b
\{g(1-R^2)+1-Q^2\}^{-(n+1)/2}\,dg}
{\int_0^\infty g^b(1+g)
\{g(1-R^2)+1-Q^2\}^{-(n+1)/2}\,dg} \\
&=& \biggl( 1+\frac{\int_0^\infty g^{b+1}
\{g(1-R^2)+1-Q^2\}^{-(n+1)/2}\,dg}
{\int_0^\infty g^{b}
\{g(1-R^2)+1-Q^2\}^{-(n+1)/2}\,dg}\biggr)^{-1} \\
&=& \biggl( 1+\frac{1-Q^2}{1-R^2} \frac{B(q/2+a+1,b+2)}{B(q/2+a+2,b+1)}
\biggr)^{-1} \\
&=& \biggl( 1+\frac{1-Q^2}{1-R^2} \frac{(n-q-3)/2-a}{q/2+a+1}
\biggr)^{-1} .
\end{eqnarray*}

On the other hand, when $q \geq n-1$, it follows that $R^2=1$, $r = n-1$,
$L(\bbm{y}|g)=(1+g)(1-Q^2)^{-(n+1)/2}$ and, hence,
%
%
\begin{equation}\label{expect-g}
H(\bbm{y}) = \frac{\int_0^\infty p(g)\,dg}{\int_0^\infty(1+g)p(g)\,dg}
=\{1+E[g]\}^{-1}.
\end{equation}

\section{\texorpdfstring{Proof of Theorem \lowercase{\protect\ref{thmconsistency}}}{Proof of Theorem 6.1}}\label{appB}

\subsection{Some preliminary lemmas}
Under the assumptions \ref{AS1} and \ref{AS2} in Section \ref
{secconsistency},
we will give the following lemmas
(Lemma \ref{lemmaprel-1} on $\bbm{X}_T$ and $\bbm{X}_\gamma$
and Lemmas \ref{lemmaprel-2}, \ref{lemmaprel-3} on $R_T^2$ and
$R_\gamma^2$)
for our main proof.
See also \citet{Fernandez-Ley-Steel-2001} and \citet{Liang-etal-2008}.
Note that \ref{AS2} implies that, for any model~$\mathcal{M}_\gamma
$, there exists a
positive definite matrix $ \bbm{H}_\gamma$ such that
%
%
\begin{equation} \label{ass-1}
\lim_{n \to\infty} \frac{1}{n}\bbm{X}'_\gamma\bbm{X}_\gamma=\bbm
{H}_\gamma.
\end{equation}
%
%
\begin{lemma} \label{lemmaprel-1}
\textup{(1)} Let $ d_1[\gamma] $ and $ d_q[\gamma] $ be
the maximum and minimum of singular values of $\bbm{X}_\gamma$.
Then $ \{d_1[\gamma]\}^2/n $ and $ \{d_q[\gamma]\}^2/n $ approach the
maximum and minimum eigenvalues of $\bbm{H}_\gamma$,
respectively.\vspace*{-5pt}

{\renewcommand\thelonglist{(2)}
\renewcommand\labellonglist{\thelonglist}
\begin{longlist}
\item\label{lemmaprel-1-2} The $q_T\times q_T$ limit
%
%
\begin{equation}
\lim_{n\to\infty} n^{-1}\bbm{X}'_T\bbm{X}_\gamma
(\bbm{X}'_\gamma\bbm{X}_\gamma)^{-1}\bbm{X}'_\gamma\bbm{X}_T = \bbm
{H}(T,\gamma)
\end{equation}
exists.\vspace*{-5pt}
\end{longlist}}

{\renewcommand\thelonglist{(3)}
\renewcommand\labellonglist{\thelonglist}
\begin{longlist}
\item\label{lemmaprel-1-3} When $\gamma\nsupseteq T$,
the rank of $\bbm{H}_T-\bbm{H}(T,\gamma)$ is given
by the number of nonoverlapping predictors and
$ \bbmm{\beta}'_T\bbm{H}_T \bbmm{\beta}_T > \bbmm{\beta}'_T \bbm
{H}(T,\gamma)\bbmm{\beta}_T $.\vspace*{-5pt}
\end{longlist}}

{\renewcommand\thelonglist{(4)}
\renewcommand\labellonglist{\thelonglist}
\begin{longlist}
\item\label{lemmaprel-1-4} $\bbm{H}_T-\bbm{H}(T,\gamma)= \bbm{0}$ for $\gamma\supsetneq T$.
\end{longlist}}
\end{lemma}
%
%
\begin{lemma}\label{lemmaprel-2}
Let $\gamma\nsupseteq T$. Then
%
%
\begin{equation}\label{plim-rss-2}
\plim_{n\to\infty}R_\gamma^2 \\
= \frac{\bbmm{\beta}'_T\bbm{H}(\gamma,T) \bbmm{\beta}_T}
{\sigma^2+\bbmm{\beta}'_T\bbm{H}_T\bbmm{\beta}_T}
\biggl(\mbox{$<$} \frac{\bbmm{\beta}'_T\bbm{H}_T\bbmm{\beta}_T}{\sigma^2+
\bbmm{\beta}'_T\bbm{H}_T\bbmm{\beta}_T}
\biggr).
\end{equation}
\end{lemma}
\begin{pf}
For the submodel $\mathcal{M}_\gamma$,
$1-R_\gamma^2$ is given by
\[
\|\bbm{Q}_\gamma(\bbm{y}-\bar{y}\bbm{1}_n)\|^2/\|\bbm{y}-\bar{y}\bbm
{1}_n\|^2
\]
with
$\bbm{Q}_\gamma=\bbm{I}-\bbm{X}_\gamma(\bbm{X}'_\gamma\bbm{X}_\gamma
)^{-1}\bbm{X}'_\gamma$.
The numerator and denominator are rewritten as
%
%
\begin{eqnarray}\label{numerator-R2}
\|\bbm{Q}_\gamma(\bbm{y}-\bar{y}\bbm{1}_n)\|^2 &=&
\|\bbm{Q}_\gamma\bbm{X}_T\bbmm{\beta}_T+\bbm{Q}_\gamma\check{
\bbmm{\varepsilon}}\|^2 \nonumber\\[-8pt]\\[-8pt]
&=& \bbmm{\beta}'_T\bbm{X}'_T\bbm{Q}_\gamma\bbm{X}_T\bbmm{\beta}_T
+ 2\bbmm{\beta}'_T\bbm{X}'_T\bbm{Q}_\gamma\bbmm{\varepsilon}+
\check{\bbmm{\varepsilon}}'\bbm{Q}_\gamma\check{\bbmm{\varepsilon
},}\nonumber
\end{eqnarray}
where $\check{\bbmm{\varepsilon}}=\bbmm{\varepsilon}-\bar{\varepsilon
}\bbm
{1}_n$ and, similarly,
\[
\|\bbm{y}-\bar{y}\bbm{1}_n\|^2= \bbmm{\beta}'_T\bbm{X}'_T \bbm{X}_T
\bbmm{\beta}_T
+ 2\bbmm{\beta}'_T\bbm{X}'_T\bbmm{\varepsilon}+\|\check{\bbmm
{\varepsilon}}\|^2.
\]
Hence, $1-R^2_\gamma$ can be rewritten as
%
%
\begin{equation} \label{1-R2}
\frac{\bbmm{\beta}'_T\{\bbm{X}'_T\bbm{Q}_\gamma\bbm{X}_T/n\}
\bbmm{\beta}_T
+ 2\bbmm{\beta}'_T\{\bbm{X}'_T\bbm{Q}_\gamma\bbmm{\varepsilon}/n\}
+ \|\bbm{Q}_\gamma\check{\bbmm{\varepsilon}}\|^2/n}
{\bbmm{\beta}'_T\{\bbm{X}'_T \bbm{X}_T/n\}\bbmm{\beta}_T
+ 2\bbmm{\beta}'_T\{\bbm{X}'_T\bbmm{\varepsilon}/n\}+\|\check{
\bbmm{\varepsilon}}\|^2/n}.
\end{equation}

In (\ref{1-R2}),
$\bbmm{\beta}'_T\bbm{X}'_T\bbmm{\varepsilon}/n $ approaches $0$ in probability
because $E[\bbmm{\varepsilon}]=\bbm{0}$,
$\operatorname{var}[\bbmm{\varepsilon}]=\sigma^2\bbm{I}_n$,
$E[\bbm{X}'_T\bbmm{\varepsilon}/n]=\bbm{0}$ and
%
%
\begin{equation} \label{var-in-probability}
\operatorname{var}(\bbm{X}'_T\bbmm{\varepsilon}/n)=
n^{-1}\sigma^2\{\bbm{X}'_T\bbm{X}_T/n\} \to\bbm{0}.
\end{equation}
Similarly $\bbmm{\beta}'_T\{\bbm{X}'_T\bbm{Q}_\gamma\bbmm{\varepsilon
}/n\} \to0 $ in probability.
Further, both $ \|\check{\bbmm{\varepsilon}}\|^2/n$ and\break
$\|\bbm{Q}_\gamma\check{\bbmm{\varepsilon}}\|^2/n$
for any $\gamma$ converge to $\sigma^2$ in probability.

Therefore, by parts \ref{lemmaprel-1-2} and \ref{lemmaprel-1-3} of
Lemma \ref{lemmaprel-1}, $R_\gamma^2$ for $\gamma\nsupseteq T$ approaches
\[
\frac{\bbmm{\beta}'_T\bbm{H}(\gamma,T) \bbmm{\beta}_T}
{\sigma^2+\bbmm{\beta}'_T\bbm{H}_T\bbmm{\beta}_T}
\biggl(\mbox{$<$} \frac{\bbmm{\beta}'_T\bbm{H}_T\bbmm{\beta}_T}
{\sigma^2+\bbmm{\beta}'_T\bbm{H}_T\bbmm{\beta}_T}\biggr)
\]
in probability.
\end{pf}
%
%
\begin{lemma}\label{lemmaprel-3}
Let $\gamma\supsetneq T$. Then:
{\renewcommand\thelonglist{(\arabic{longlist})}
\renewcommand\labellonglist{\thelonglist}
\begin{longlist}
\item\label{lemmaprel-3-1} $ R_\gamma^2 \geq R^2_T$ for any $n$ and
%
%
\begin{equation}\label{plim-rss-1}
\plim_{n\to\infty}R_T^2=\plim_{n\to\infty}R_\gamma^2
=\frac{\bbmm{\beta}'_T\bbm{H}_T\bbmm{\beta}_T}{\sigma^2+\bbmm{\beta
}'_T\bbm{H}_T\bbmm{\beta}_T}.
\end{equation}
\item\label{lemmaprel-3-2}
$\{(1-R_T^2)/(1-R_\gamma^2)\}^n $ is bounded from above in probability.
\end{longlist}}
\end{lemma}
\begin{pf}
(1)
When $\gamma\supsetneq T$,
$\bbm{Q}_\gamma\bbm{X}_T=\bbm{0}$.
Hence, as in (\ref{1-R2}), we have
%
%
\begin{eqnarray}\label{1-R2-proper-sub}
1-R_\gamma^2
&=&\frac{\|\bbm{Q}_\gamma\check{\bbmm{\varepsilon}}\|^2/n}
{\bbmm{\beta}'_T\{\bbm{X}'_T \bbm{X}_T/n\}\bbmm{\beta}_T
+ 2\bbmm{\beta}'_T\{\bbm{X}'_T\bbmm{\varepsilon}/n\}+
\|\check{\bbmm{\varepsilon}}\|^2/n}, \nonumber\\[-9pt]\\[-9pt]
1-R_T^2
&=&\frac{\|\bbm{Q}_T \check{\bbmm{\varepsilon}}\|^2/n}
{\bbmm{\beta}'_T\{\bbm{X}'_T \bbm{X}_T/n\}\bbmm{\beta}_T
+ 2\bbmm{\beta}'_T\{\bbm{X}'_T\bbmm{\varepsilon}/n\}+
\|\check{\bbmm{\varepsilon}}\|^2/n}.
\nonumber
\end{eqnarray}
Since
$ \|\bbm{Q}_T \check{\bbmm{\varepsilon}}\|^2/n>
\|\bbm{Q}_\gamma\check{\bbmm{\varepsilon}}\|^2/n$
for any $n$ and both approach $\sigma^2$ in probability,
part~\ref{lemmaprel-3-1} follows.

(2) By (\ref{1-R2-proper-sub}),
$ (1-R^2_T)/(1-R_\gamma^2)$ is given by
$ \|\bbm{Q}_T\check{\bbmm{\varepsilon}}\|^2/\|\bbm{Q}_\gamma\check{
\bbmm{\varepsilon}}\|^2$.
Further, we have
\[
1 \leq\frac{1-R^2_T}{1-R_\gamma^2}=
\frac{\|\bbm{Q}_T\check{\bbmm{\varepsilon}}\|^2}{\|\bbm{Q}_\gamma\check
{\bbmm{\varepsilon}}\|^2}
\leq
\frac{\|\check{\bbmm{\varepsilon}}\|^2}{\|\bbm{Q}_\gamma\check{
\bbmm{\varepsilon}}\|^2}
=\frac{1}{W_\gamma},
\]
where $W_\gamma\sim(1+\chi^2_{q_\gamma}/\chi^2_{n-q_\gamma
-1})^{-1} $, for independent $\chi^2_{n-q_\gamma-1}$ and $\chi
^2_{q_\gamma}$.
Hence,
\begin{eqnarray*}
\{ 1+\chi^2_{q_\gamma}/\chi^2_{n-q_\gamma-1}\}^{-n}
&=& \bigl\{ 1+\{n/\chi^2_{n-q_\gamma-1}\}
\{\chi^2_{q_\gamma}/n\} \bigr\}^{-n} \\[-2pt]
&\sim& \exp(-\chi^2_{q_\gamma}) \qquad\mbox{as }n \to\infty
\end{eqnarray*}
since $ \chi^2_{n-q_\gamma-1}/n \to1$ in probability.
Therefore, $ W_\gamma^{-n}$ is bounded in probability from above
and part \ref{lemmaprel-3-2} follows.
\end{pf}

\subsection{\texorpdfstring{The proof of Theorem \protect\ref{thmconsistency}}{The proof of Theorem 6.1}}
Note that
\[
\nu_1^{-1} \leq1-Q^2_\gamma\leq1
\]
by (\ref{R2Q2}),
\[
\nu_1^{-q/2} \leq\prod_{i=1}^q \nu_i^{-1/2} \leq1,
\]
because the $\nu_i$'s are descending,
\[
\frac{B(q/2+a+1,(n-q-3)/2-a)}{B(a+1,(n-q-3)/2-a)}=
\frac{\Gamma(q/2+a+1)}{\Gamma(a+1)}\frac{\Gamma(\{n-q-1\}
/2)}{\Gamma(\{n-1\}/2)}
\]
and
\[
\lim_{n \to\infty}(n/2)^{q/2} \frac{\Gamma(\{n-q-1\}/2)}{\Gamma(\{
n-1\}/2)}=1\vadjust{\goodbreak}
\]
by Stirling's formula.
Then, by (\ref{BC-2}), there exist $c_1(\gamma) < c_2(\gamma)$
(which do not depend on $n$)
such that
\[
c_1(\gamma) < \{n^{q_\gamma}(1-R^2_\gamma)^{n}\}^{1/2}
\frac{\BF\bgvn(a,\nu)}
{(1-R^2_\gamma)^{(q_\gamma+3)/2+a}}
<c_2(\gamma)
\]
for sufficiently large $n$.
By Lemmas \ref{lemmaprel-2} and \ref{lemmaprel-3}, $R^2_\gamma$ goes
to some constant
in probability.
Hence, to show consistency, it suffices to show that
%
%
\begin{equation} \label{main-plim}
\plim_{n \to\infty}
n^{q_T-q_{\gamma}}
\biggl(\frac{1-R^2_T}{1-R^2_{\gamma}}\biggr)^{n}
=0.
\end{equation}

Consider the following two situations:
\begin{longlist}[(2)]
\item[(1)]
$\gamma\nsupseteq T$:
by Lemmas \ref{lemmaprel-2} and \ref{lemmaprel-3},
$ (1-R^2_T)/(1-R^2_{\gamma})$
is strictly less than~$1$ in probability.
Hence, $ \{(1-R^2_T)/(1-R^2_{\gamma})\}^n$
converges to zero in probability
exponentially fast with respect to $n$.
Therefore, no matter what value $ q_T-q_{\gamma}$ takes,
(\ref{main-plim}) is satisfied.
\item[(2)] $\gamma\supsetneq T$:
by Lemma \ref{lemmaprel-3},
$\{(1-R^2_T)/(1-R^2_{\gamma})\}^{n}$ is bounded in probability.
Since $ q_{\gamma}> q_T$, (\ref{main-plim}) is satisfied.
\end{longlist}
\end{appendix}

\section*{Acknowledgments}

We are very grateful to a referee for wonderful insights which
substantially helped us to strengthen this paper.


%

\printaddresses

\end{document}